\documentclass[12pt,a4paper]{article}
\usepackage{jheppub}

 \usepackage{graphicx}
  \usepackage{epstopdf}

\newcommand{\bej}[1]{ \begin{equation}\label{#1} }
\newcommand{\eej}{\end{equation}}
\newcommand{\beaj}[1]{\begin{eqnarray}\label{#1} }
\newcommand{\eeaj}{\end{eqnarray}}
\newcommand{\eq}[1]{(\ref{#1})}
\def\ZZZ{{\hskip-3pt\hbox{ Z\kern-1.6mm Z}}}
\def\zzz{{\hskip-3pt\hbox{ z\kern-1mm z}}}

\newcommand{\tb}{\tilde b}

\newcommand{\bd}{\bar{\rm D}}

\newcommand{\N}{\frac{m_{2}}{k_{2}}-\frac{m_{1}}{k_{1}}}

\newcommand{\be}{\begin{equation}}
\newcommand{\ee}{\end{equation}}
\newcommand{\ben}{\begin{eqnarray}\displaystyle}
\newcommand{\een}{\end{eqnarray}}

\newcommand{\p}{\partial}

\def\one{{\hbox{ 1\kern-.8mm l}}}
\def\zero{{\hbox{ 0\kern-1.5mm 0}}}

\def\be{\begin{equation}}       
\def\ee{\end{equation}}         

\def\bea{\begin{eqnarray}}      
\def\eea{\end{eqnarray}}
                                
\def\ba{\begin{array}}
\def\ea{\end{array}}
\def\bd{\begin{displaymath}}
\def\ed{\end{displaymath}}

\def\eq{\begin{equation}}
\def\eqe{\end{equation}}
\def\eqa{\begin{eqnarray}}
\def\eqae{\end{eqnarray}}
\def\ena{\end{eqnarray}}

\def\nn{\nonumber}

\def\unit{1 \hskip-.3em \raise2pt\hbox{$ \scriptstyle |$ } }

\def\a{\alpha}
\def\b{\beta}

\def\e{\epsilon}           
   
\def\g{\gamma}

\def\l{\lambda}

\def\n{\nu}
  \def\w{\omega}
\def\p{\pi}                
\def\r{\rho}                                     
\def\s{\sigma}                                   
\def\t{\tau}

\def\D{\Delta}

\def\cs{{\cal S}}

\def\bd{\begin{displaymath}}
\def\ed{\end{displaymath}}

\def\6{\partial}
\def\N4{{\cal N}=4}

\def\bop#1{\setbox0=\hbox{$#1M$}\mkern1.5mu
        \vbox{\hrule height0pt depth.04\ht0
        \hbox{\vrule width.04\ht0 height.9\ht0 \kern.9\ht0
        \vrule width.04\ht0}\hrule height.04\ht0}\mkern1.5mu}

\def\>{\rangle} 

\def\<{\langle} 
\def\Dsl{D \hskip-.6em \raise1pt\hbox{$ / $ } }

\def\+{\oplus}

\title{Spinning strings and minimal surfaces in $AdS_3$ with mixed 3-form fluxes }

\author{Justin R. David and } 
\author{Abhishake Sadhukhan}

\affiliation{ Centre for High Energy Physics,\\
Indian Institute of Science,\\ C.V. Raman Avenue, Bangalore 560012, India}

\emailAdd{justin, abhishake@cts.iisc.ernet.in}

\abstract{
Motivated by the recent proposal for the S-matrix in $AdS_3\times S^3$ with mixed 
three form fluxes,
 we  study  classical folded string spinning in $AdS_3$ with  both Ramond   and Neveu-Schwarz   three form fluxes. 
 We solve the equations of motion of these  strings and obtain their dispersion relation 
 to the leading order 
 in  the Neveu-Schwarz flux $b$. 
 We show that  dispersion relation for the spinning strings  with
 large spin ${\cal S}$  acquires a term
 given by $-\frac{\sqrt{\l}}{2\pi} b^2\log^2 {\cal S}$ in addition to the usual $\frac{\sqrt\lambda}{\pi} \log {\cal S}$ term
 where  
 $\sqrt{\lambda}$ is proportional to the square of the radius of $AdS_3$.  
  Using $SO(2,2)$ transformations and re-parmetrizations we show that these 
 spinning strings can be related to light like Wilson loops in $AdS_3$ with Neveu-Schwarz
 flux $b$. We observe that the logarithmic divergence in the
 area of the light like Wilson loop is also deformed 
 by precisely the same  coefficient of the $  b^2 \log^2 {\cal S}$  term in the dispersion relation of the spinning 
 string. 
 This result indicates that the coefficient of $ b^2 \log^2 {\cal S}$ has a property similar to
 the coefficient of the $\log {\cal S}$ term, known as cusp-anomalous dimension, 
  and can  possibly be determined to all 
 orders in the coupling $\lambda$ using the recent proposal for  the S-matrix. }

\begin{document}
\maketitle
 \section{Introduction}

Classical string solutions propagating in various $AdS\times M$ 
background  have played an important role in the study of the $AdS/CFT$ correspondence. 
The anomalous dimensions of various operators of the field 
theory with large charges can be obtained
by examining the dispersion of classical strings. 
Non-local operators like Wilson loops are  also  represented  as minimal surfaces in the 
dual geometry. 
One particular classical solution which has been crucial in the detailed study of 
${\cal N}=4$ Yang-Mills is the folded spinning string solution in $AdS_5$. 
This solution was originally found by \cite{Gubser:2002tv} and was studied in more detail 
in \cite{Frolov:2002av}. 
The dispersion relation of the  string 
which effectively moves in a $AdS_3$ subspace of $AdS_5$  with large spin $\cs$ is 
given by 
\be \label{dispads5}
\Delta=\cs+\frac{\sqrt{\l}}{\pi}\log \frac{\cs}{\sqrt{\l}},\quad \l\rightarrow \infty. 
\ee
Here $\Delta$ is the energy of the string. 
The spinning folded string  is dual to twist two operators of the form ${\rm Tr }(\Phi \partial^S \Phi )$,
where the $\Phi$ is one of the adjoint scalars in ${\cal N}=4$ Yang-Mills and $\partial$ indicated
spatial derivatives.
 From a perturbative analysis in the field theory it can be shown that in the planar limit 
 the anomalous dimensions of these operators with large $\cs$ obey the relation
 \be 
\D-(\cs+2)=f(\l) \log \cs + O(1/\cs). 
\ee
The function $f(\lambda)$ at weak t'Hooft coupling is given by $f(\lambda) = \lambda$ 
while the behaviour of $f(\lambda)$ at
 strong coupling  can be read out from the dispersion relation of the 
spinning string in (\ref{dispads5}). 
$f(\lambda)$ is related to a variety of different physical observables \cite{Correa:2012at}.
One particular relation  which is of interest in this paper is that $f(\lambda)$ 
determines the expectation value of the  Wilson loop operator which 
has a cusp in its contour. 
It can be shown entirely from a field theory analysis \cite{Korchemsky:1988si, Korchemsky:1992xv}
 that $f(\lambda)$ determines the
logarithmic divergence of a Wilson loop which makes a turn of angle $\gamma$ from the
straight line.   In the  limit of large cusp angle the Wilson loop is given by 
\be
W=\left(\frac{L}{\e} \right)^{-f(\l)\,\g }, 
\ee
where $L$  and $\epsilon$ are the ultra-violet and Infra-red cut off respectively. 
The function  $f(\lambda)$ is called the cusp anomalous dimensions in literature.
The area of the minimal surface corresponding 
to the cusp Wilson loop in the dual geometry  exhibits a  similar logarithmic divergence 
and $f(\lambda) = \frac{\sqrt{\lambda}} {\pi} $ \cite{Kruczenski:2002fb, Makeenko:2002qe}. 
The minimal surface also lies only in a $AdS_3$ sub-space of $AdS_5$. 
Indeed the classical spinning string solution after a series of conformal transformations
and re-parametrizations can be related to the minimal surface solution \cite{Kruczenski:2007cy}. 
The function $f(\lambda)$ is that it has been 
determined to all orders in $\lambda$  by using integrability \cite{Beisert:2006ez}. 

Another example of  holographic dual pair is that the case of 
strings on $AdS_3\times S^3 \times T^4$ and  the ${\cal N} = (4, 4)$ super 
conformal field theory associated with the D1-D5 system. 
Motivated by studying integrability of the string theory in this background
semi-classical solutions like magnons and the folded spinning strings have also been 
studied in this background \cite{David:2008yk,David:2010yg,Abbott:2012dd,Beccaria:2012kb,
Beccaria:2012pm,Abbott:2013ixa,Rughoonauth:2012qd,Sundin:2013ypa,Sundin:2014sfa,Sundin:2013uca}. Recently it has been shown that the background $AdS_3\times S^3$ is supported by 
both Ramond and Neveu-Schwarz three form fluxes, the string theory is integrable 
\cite{Cagnazzo:2012se, Wulff:2014kja}. 
There is a proposal for the S-matrix in this background \cite{Hoare:2013pma,Hoare:2013ida,Bianchi:2014rfa}.
The giant magnon solution  and the finite gap equations has been studied in this background by
\cite{Hoare:2013lja} and \cite{Babichenko:2014yaa} respectively. Short string solutions in presence of Neveu Schwarz B field in $AdS_5\times S^5$ geometry were studied in \cite{Rashkov:2002zt}. Our goal in this paper is to study the behaviour of the spinning folded  string 
solution in the background of $AdS_3\times S^3$ with both Ramond as
well as Neveu-Schwarz three form fluxes.  This study is motivated from the 
recent proposal of the S-matrix for this background.

Once the Neveu-Schwarz three form flux is turned on, the equations of motion
of the string are affected 
and therefore the folded spinning string solution cannot be simply embedded in 
$AdS_3$.   Earlier studies  of classical strings 
in $AdS_3$ with Neveu-Schwarz fluxes are
 \cite{Maldacena:2000hw,Loewy:2002gf, Lee:2008sk}.  In this paper we 
we solve the equations of motion of the classical folded spinning string solution in
$AdS_3$ in terms of a perturbative expansion in the Neveu-Schwarz flux $b$. 
We show that the dispersion relation of the spinning strings in the 
large $\cs$ limit  is given by
\begin{equation} \label{dispn}
\Delta=\cs+\frac{\sqrt{\l}}{\pi}\log \cs  -  \frac{\sqrt{\lambda} }{2\pi} b^2 
\log^2\cs, 
\end{equation}
where $\lambda$ is related to the radius of $AdS_3$ and we have retained the leading 
terms in the large $\cs$ limit. 
The order of perturbative expansion 
which results in  the dispersion relation (\ref{dispn})  is the following. 
We first perform a perturbative expansion in $b$. At each  order in $b$ we 
retain the leading term in the large $\cs$ limit.  We also require 
$b \log\cs <<1$ so that the expansion in (\ref{dispn}) makes sense. 
$\log^2 \cs$ terms are known to occur in the anomalous dimensions of 
twist two operators in 4 dimensional theories with lesser super symmetries compared
to ${\cal N}=4$ Yang-Mills \cite{Basso:2006nk}.  
But these terms are always suppressed by $1/\cs$ and therefore 
are not relevant in the large $\cs$ limit. 

From (\ref{dispn})  we note that at order $b^2$ the leading term is 
proportional to $\log^2\cs$.  
As we have discussed earlier, the  coefficient of the  $\log\cs$ term in the dispersion relation of the 
spinning string is also the coefficient of the logarithmic divergence seen in the 
area of the minimal surface corresponding to the cusp Wilson line. 
When these classical solutions are embedded in $AdS_5$, the reason 
that these coefficients must agree has a purely perturbative 
field theoretic explanation \cite{Korchemsky:1988si, Korchemsky:1992xv}.   
It was argued in \cite{Kruczenski:2007cy} from the dual gravity side the reason 
that the coefficient of the $\log\cs$ term agreed with the cusp anomalous dimension 
is that the classical solutions can be related to each other by $SO(2, 2)$ transformations. 
In fact in \cite{Kruczenski:2007cy}, the one loop corrections around both the spinning string and the 
cusp Wilson line were evaluated and  shown to agree. 
In light of this fact it is interesting to study the  area of the minimal surface in presence of the
Neveu-Schwarz field. 
We show that the equations of motion for the minimal surface and its action
 in presence of the Neveu-Schwarz field can be solved exactly.  
 We show that the logarithmic divergence of the 
 modulus of the expectation value of Wilson loop is determined by the function 
 \begin{equation}
 f(\lambda) = \frac{\sqrt\lambda}{\pi} - \frac{\sqrt{\lambda} }{2\pi} b^2 + O(b^4).  
 \end{equation}
 We observe that the  coefficient of $b^2$ is  identical to the coefficient of the $b^2\log^2\cs$ in the 
 dispersion relation of the spinning string. 
 To argue   that  there is a relation of the coefficient of $b^2$ in the logarithmic divergence of the 
 area of the minimal surface corresponding to the Wilson loop  to 
 the coefficient of $b^2\log^2\cs$  in the dispersion relation of the spinning string, 
 we note the following. 
 \begin{enumerate}
 \item
We first relate  the scaling limit of the spinning 
 string solution found to $O(b^2)$ to the minimal surface corresponding to the 
 cusp Wilson line by performing a series of $SO(2,2)$ 
 transformations\footnote{Note that the world sheet sigma model in the presence of 
 both Ramond and Neveu-Schwarz fields has global $SO(2,2)$ symmetry.} and 
 re-parametrizations as done by  \cite{Kruczenski:2007cy}  in the absence of
 the Neveu-Schwarz field.
 \item
 We also show that it is only the 
 logarithmic divergence  in the area of the minimal surface that is universal.  There 
 is a class of minimal surfaces which ends on the cusp Wilson line and 
 the logarithmic divergence in the area remains the same but admit other divergences
 which depends on the parameters of the minimal surface.  
 \end{enumerate}
 Using these facts,  if we extrapolate the observation of \cite{Kruczenski:2007cy}   to $b\neq 0$ 
 it is natural to  compare the coefficient of the $b^2\log^2\cs$  to the coefficient in the 
 logarithmic divergence of the area of the minimal surface. 
 The observation that these coefficients are same 
  suggests that the coefficient of $b^2\log^2\cs$  has a property 
 similar to that of the $\log\cs$ term in the dispersion relation of the spinning string. 
 It will be interesting if the S-matrix proposal of  \cite{Hoare:2013pma,Hoare:2013ida} can be used to derive this dispersion 
 relation and whether one can determine the $b^2$ term to all orders in $\lambda$.

 This paper is organized as follows. 
 In section 2 after a brief review of the supergravity background with both Neveu-Schwarz and 
 Ramond 3-form fluxes in $AdS_3$ we solve the equations of motion for the spinning string 
 in $AdS_3$ with angular momentum in the $S^3$ to the leading order in the Neveu-Schwarz
 flux $b$. We show that in the large spin limit the dispersion relation is given by 
 (\ref{dispn}).  We will  again derive this dispersion relation in the scaling limit of the 
 long string. 
 As a check on our perturbation theory, we also obtain the dispersion relation 
 for small strings and show that we obtain the BMN dispersion relation. 
 In section 3 we first derive the minimal surface corresponding to the cusp
 Wilson line in the background with 
 mixed  3-form flux and evaluate its area exactly. 
 We show that the coefficient of $b^2$ in the area is precisely the same as the coefficient
 of $b^2 \log ^2\cs$ term in the dispersion relation (\ref{dispn}). 
 We then relate the scaling limit of the spinning string solution to that of 
cusp Wilson line using  the $SO(2,2)$  symmetry of the sigma model and 
re-parametrization invariance. Section 4 contains our conclusions.

\section{Strings in $AdS_3\times S_3$ with mixed form fluxes}

To set up our notations and conventions we first write down 
the background solution which we will work in. 
It is a solution of the type IIB action with $AdS_3\times S^3 \times M^4$ geometry. 
$M^4$ can be $T^4$ or $K^3$. This will not be relevant for the discussions in this 
paper.  The solution has RR and NS-NS fluxes along the $AdS_3$ and $S^3$ directions. 
The type IIB background fields which are turned on in this solution are
\begin{eqnarray}\label{metrics}
ds^2 &=&  ds_{AdS_3}^2 + ds_{S^3}^2 +  ds_{T^4}^2, \\ \nn
ds_{AdS_3}^2&=&-\cosh^2{\rho} dt^2+d\rho^2+\sinh^2{\rho} d\phi^2, \\ \nn
ds_{S^3}^2&=&d\b_1^2+\cos^2{\b_1}(d\b_2^2+\cos^2{\b_2}d\b_3^2). 
\end{eqnarray}
Here we have assumed that the compact manifold to be the torus $T^4$ for definiteness. 
The NS-NS and RR 3-form fluxes are given by 
\begin{eqnarray}\label{flux}
H_{t\rho\phi}^{(3)}&=&-2 b \cosh \rho \sinh \rho,\quad F_{t\rho\phi}^{(3)}=-2\sqrt{1-b^2} \cosh \rho \sinh \rho ,\quad\\ \nn
 H_{\b_1 \b_2\b_3}^{(3)}&=&-2 b \cos^2 \b_1\cos \b_2 \quad F_{\b_1 \b_2\b_3}^{(3)}=-2\sqrt{1-b^2}\cos^2 \b_1\cos \b_2, \nn
\end{eqnarray}
where $0\leq b\leq 1$. 
As the parameter $b$ is tuned form $0$ to $1$, the solution 
interpolates from purely RR 3-from flux to purely NS-NS flux. 
All the remaining fluxes are set to zero,  the dilaton, $\Phi$ is constant and can be  set to zero.  
We have taken the radius of $AdS_3$ and $S^3$ to be unity.  
We will incorporate the radius of $AdS_3$ in the sigma model coupling. 
This background is the solution of type IIB equations of motion in the 
Einstein frame.  For completeness  and as a check of 
our normalizations we write down the equations of motion for the metric
\begin{eqnarray}\label{e2b}
&&R_{MN}\,-\, \frac{1}{2}G_{MN}R\,=\,\frac{1}{2} \Big ( \partial_M\Phi\partial_N\Phi\,-\,\frac{1}{2}G_{MN}\partial_P\Phi \partial^P\Phi \Big )\, \\ \nonumber
& & +\,\frac{1}{2}e^{2\Phi} \Big ( F^{(1)}_M F^{(1)}_N\,-\,\frac{1}{2}G_{MN}F_1^2 \Big )\,+\frac{1}{2.3!}e^{\Phi} \Big ( 3F^{(3)}_{MPQ}F^{(3)PQ}_N\,-\, \frac{1}{2}G_{MN}F_3^2 \Big )\,
\\ \nonumber
& & 
+\,\frac{5}{4.5!}F^{(5)}_{MPQRS}F^{(5)PQRS}_N\,+\,+\frac{1}{2.3!}e^{-\Phi}\Big ( 3H^{(3)}_{MPQ}H^{(3)PQ}_N\,-\, \frac{1}{2}G_{MN}H_3^2 \Big) \,\, . \\ \nn
\end{eqnarray}
The background metric in (\ref{metrics}) and the fluxes (\ref{flux}) can be easily shown to satisfy the equation 
(\ref{e2b}) with the dilaton set to zero.

\subsection{Classical solution for spinning strings in  $AdS_3\times S_3$}

To begin deriving the classical solution of strings in $AdS_3\times S_3$ with mixed 
3-form fluxes we first write down the sigma model in the background given by 
(\ref{metrics}) and (\ref{flux}). 
The Polyakov action is given by 
\be\label{spol}
S_{pol}=\frac{\sqrt{\l}}{4\pi}\int d\tau d\s \Big[\sqrt{-h}h^{ab}G_{mn}\partial_a X^m \partial_b X^n-\e^{ab} B_{mn}\partial_a X^m \partial_b X^n\Big]. 
\ee
where $h^{ab}$ in conformal  gauge is  given by $h^{ab}=\mbox{diag}(-1,1)$.   
The antisymmetric tensor  $\e^{01}=1$ and the indices $m, n$ run from $0$ to $5$. 
The directions $0, 1, 2$ label $AdS_3$ while $3, 4, 5$ label the $S^3$. 
The classical solutions of interest in this paper have no dynamics along the $T^4$, therefore
from now on we will ignore these directions. 
Note that we have re-instated the radius of $AdS_3$ and $S^3$ in the sigma model coupling
$\sqrt{\lambda}$ which is given by 
\begin{equation}
 \sqrt{\lambda} = \frac{R^2}{\alpha'}, 
\end{equation}
where $R$ is the radius of $AdS_3$ and $\alpha'$ is the string length squared. 
The interesting dynamics of the spinning string solutions we consider will take 
place in $AdS_3$,  for which we find it convenient to  work with the following global metric
\be
\label{adsr}
ds^2=-(1+r^2) d\tilde t^2+\frac{dr^2}{1+r^2}+r^2 d\tilde \phi^2. 
\ee
This metric is related to the $AdS_3$ metric given in (\ref{metrics}) by the 
 coordinate transformation $r=\sinh{\rho}$. 
 The NS-NS flux along $AdS_3$ in this co-cordinate system is given by 
\be
B_{t\phi}=b r^2. 
\ee
The metric on $S^3$ and the NS-NS flux on $S^3$ is taken to be as given in (\ref{metrics})
and (\ref{flux}) respectively. 
We choose the following ansatz for the classical solutions we consider.
\be \label{anst}
\tilde t=c_1\t+t(\s),\quad \tilde \phi=c_2\t+\phi(\s),\quad r=r(\s),\quad \b_1=\b_2=0,\quad \b_3=\w \t.
\ee
Here $\tilde \phi$ is the angle in $AdS_3$ and $\tilde t$ is the global time while $r$ is the radial direction.
We look for solutions which satisfy the condition
\begin{equation}\label{closed}
t(\sigma+ 2\pi ) = t(\sigma),  \qquad
r(\sigma + 2\pi) = r(\sigma). 
\end{equation}
Just as in the giant magnon solution of \cite{Hofman:2006xt} we do not impose periodic boundary 
conditions on the co-ordinate $\phi$. We will however show that 
a closed string solution can be constructed by considering several periods in the 
world sheet $\sigma$ direction. 
This ansatz in (\ref{anst}) is a generalization of the  folded spinning string solution in the absence of 
NS-NS flux studied in \cite{Frolov:2002av}.  
 The functions $t(\sigma)$ and $\phi(\sigma)$  vanish when  $b=0$. 
 The angle co-ordinate $\tilde\phi$ is  function of $\sigma$ when $b\neq 0$. The solution 
 we construct 
 at a given instance of time in the $r, \tilde \phi$ plane will not be a simple line, which
  is the signature of the folded string. Since $\tilde\phi$ is a function of $\sigma$ it will look like a
  smoothed  or a blown up version
   of the folded string, turning around smoothly in the $r, \tilde\phi$ plane. 
  We will continue to call this the folded spinning string as it reduces to the folded solution 
  when $b=0$. 
 
Substituting the ansatz into the  world sheet action (\ref{spol}), we find that it  reduces to 
\begin{eqnarray}
\label{action1}
S=\frac{\sqrt{\l}}{4\pi}\int d\tau d\s \Big[&&-(1+r^2)(-c_1^2+t'^2)+\frac{1}{1+r^2}r'^2+r^2(-c_2^2+\phi'^2)\\ \nonumber
&&-2 b r^2(c_1 \phi' - c_2 t')\Big]. 
\end{eqnarray}
The NS-NS flux in the $S^3$ direction does not contribute to the action since the 
ansatz in (\ref{anst}) does not have world sheet $\sigma$ direction in the $S^3$ directions.   
Note that the action in (\ref{action1}) does not explicitly depend on $t$ and $\phi$, therefore 
the corresponding conjugate momenta are conserved. 
This leads to the following equations 
\be
\label{tp}
t'=\frac{b c_2 r^2-k_1}{1+r^2},\qquad\qquad \phi'=\frac{b c_1r^2-k_2 }{r^2},
\ee
where $k_1, k_2$ are constants of motion. 
The Virasoro constraint  corresponding to  the 
vanishing of world sheet momentum leads to the constraint 
\be
\label{vir0} 
c_1 k_1=c_2 k_2. 
\ee
The Virasoro constraint corresponding to the vanishing of  world sheet energy leads to 
a first order differential equation for the function $r$. 
\be
\label{vir1}
r'^2r^2=A r^6+B r^4+Cr^2+D,  
\ee
where
\begin{eqnarray}
\label{ABcd}
 A&=&c_1^2+c_2^2 b^2-c_2^2-c_1^2 b^2,  \\ \nonumber
B&=&2 c_1^2-c_2^2-c_1^2 b^2-2bc_2k_1+2bc_1k_2-\w^2 , \\ \nonumber
C&=&c_1^2+k_1^2+2bc_1k_2-k_2^2-\w^2,  \\ \nonumber
D&=&-k_2^2. 
\end{eqnarray}
It can be verified that  the solutions for $t', \phi'$ and $r'$ given in  (\ref{tp}) and (\ref{vir1})
solves the second order equations of motion derived from the sigma model.

Note that the equations simplify  when $b=1$, that is the situation when there is a pure
NS-NS flux. The coefficient $A$ vanishes and the polynomial on the right hand side
of (\ref{vir1}) 
reduces to a quartic polynomial. This is the limit studied in 
\cite{Loewy:2002gf} for which exact solutions were obtained.
  We will first write down the solutions for for a general $b$ and then
set up a perturbative expansion about $b=0$. 
As we have mentioned earlier,  $t(\sigma)$ and $\phi(\sigma)$ must vanish 
when $b=0$ for the solution to reduce to the folded spinning string.
From  (\ref{tp}), this implies  that $k_1, k_2$ must vanish when $b=0$. 
Assuming there exists a well defined perturbation theory about $b=0$ 
we look for solutions with $k_1, k_2$ vanish linearly with $b$. 
Therefore $k_1, k_2$ admits an expansion given by 
\be\label{pexpand}
k_1=k_{1}^{(0)}b+k_{1}^{(1)} b^2 + \cdots ,\qquad\qquad k_2=\frac{c_1 k_1}{c_2}. 
\ee
We can now integrate the equation for $r'$. 
For this we need to find the turning points of the equation in (\ref{vir1}). 
 We see that from the expressions  for 
$A, B, C, D$  in (\ref{ABcd})  and the expansion in (\ref{pexpand}),  when $b=0$ the 
three roots of 
cubic polynomial in $r^2$ given by $Ar^6 + Br^4 + Cr^2 + D$  which determines $r'$
are $-1, 0,  \frac{c_1^2 -\omega^2}{ c_2^2 - c_1^2}$, we label these roots as 
$R_1^{(0)}, R_3^{(0)}$ and $R_2^{(0)}$ respectively. 
The turning points  for $r'$ for the folded spinning string solutions are 
at $0$ and $\frac{c_1^2 -\omega^2}{ c_2^2 - c_2^2}$, that is $R_3^{(0)}$ and $R_2^{(0)}$. 
Let $R_3$ and $R_2$ be the 
 roots continuously connected to the roots 
$R_3^{(0)}$ and $R_2^{(0)}$  respectively when $b$ is set to zero.
Let us  rewrite the equation (\ref{vir1}) as 
\be\label{vir2}
r'=\frac{\sqrt{A(r^2-R_1)(r^2-R_2)(r^2-R_3)}}{r}, 
\ee
$R_1, R_2, R_3$ are the roots continuously connected to $R_1^{(0)}, R_2^{(0)}$ and 
$R_3^{(0)}$ respectively. 
The folded string string satisfies the periodicity property  $r(\sigma+ 2\pi) = r(\sigma)$. 
This is ensured as follows. 
The interval $0\le \s < 2\pi$ is split into 4 segments.
For $0\le \s <\pi/2$, $r(\s)$ increases from $\sqrt{R_3}$ to $\sqrt{R_2}$. $r(\s)$. Then
 $r(\s)$ decreases back to $\sqrt{R_2}$ as $\s$ goes from $\pi/2$ to $\pi$. 
 Integrating the equation in  (\ref{vir2})  between 
 $\sqrt{R_3}$ to $\sqrt{R_2}$. $r(\s)$  we obtain
\be
\label{vireq}
2\pi=\int^{2\pi}_0 d\s=4\int^{\sqrt{R_2}}_{\sqrt{R_3}}\frac{r dr }{\sqrt{A (r^2-R_1)(r^2-R_2)(r^2-R_3)}}.
\ee
We can reduce the integral in (\ref{vireq}) to a known function by 
substitution $r^2=x+R_3$. The integral then becomes
\be
2\pi = \frac{4}{2\sqrt{A}}\int_0^{z_1}\frac{dx}{\sqrt{x(x-z_1)(x-z_2)}},
\ee
where $z_1=R_2-R_3$ and $z_2=R_1-R_3$. 
It is now easy to recognize that after an appropriate scaling the integral reduces to the 
hypergeometric function. Thus we obtain the condition
\be
\label{eq1}
\sqrt{Az_2}={}_2F_1(\frac{1}{2},\frac{1}{2},1,\frac{z_1}{z_2}).
\ee

Let us now examine the   difference  $\phi(\sigma + 2\pi) -\phi(\sigma) $.
After substituting the expression for $\phi'$ from (\ref{tp}) in the above equation we obtain
\begin{equation}\label{phic}
\phi( \sigma + 2\pi) - \phi(\sigma) = b c_1 2\pi-\frac{c_1 k_1}{c_2}\int_0^{2\pi} \frac{d\s }{r^2}.
\end{equation}
The  integral in  (\ref{phic}),  can be rewritten as 
\begin{eqnarray}\label{phint}
\int_0 ^{2\pi} \frac{d\s}{r^2}=\frac{4}{\sqrt{A}}\int^{\sqrt{R_2}}_{\sqrt{R_3}}\frac{dr}{r \sqrt{(r^2-R_1) (r^2-R_2) (r^2-R_3)}}.
\end{eqnarray} 
After   changing    variables to  $x=r^2-R_3$, the integral reduces to 
\begin{eqnarray}
\int_0 ^{2\pi} \frac{d\s}{r^2}&=& \frac{2}{\sqrt{A}}\int^{z_1}_0\frac{dx}{(x+R_3)\sqrt{x(x-z_1)(x-z_2)}}
\nonumber, \\
&=& \frac{4}{R_3 \sqrt{A z_2}} \Pi (u_1|v_1),
\end{eqnarray}
where $u_1=-\frac{z_1}{R_3}$, $v_1=\frac{z_1}{z_2}$. 
$\Pi(a|b)$ is the complete elliptic integral of the third kind defined by 
\be
\Pi(u|v)=\frac{1}{2}\int^1_0\frac{dy}{(1-uy)\sqrt{y(1-y)(1-vy)}}.
\ee
Therefore we obtain
\be \label{feq2}
\phi(2\pi) - \phi(0) = 2\pi b c_1-\frac{4 k_2}{R_3 \sqrt{z_2}}\Pi(u_1|v_1). 
\ee 
In general this difference will not vanish, but after sufficiently large number
of periods in the world sheet $\sigma$ direction we can ensure that the 
string closes. This will be shown explicitly in the scaling limit \ref{scaling}. 
Finally we examine the closed string condition $t(\sigma + 2\pi) = t(\sigma)$. 
This can be written as 
\begin{equation}
\int_0^{2\pi} d\sigma t' =0.
\end{equation}
Substituting the value of $t'$ from (\ref{tp}) and using similar change of variables, the 
above condition can be written again in terms of elliptic function of the third kind. 
This results in the following condition
\be \label{feq3}
2\pi b c_2-\frac{4(b c_2+k_1)}{(1+R_3)\sqrt{Az_2}}\Pi(u_2|v_2)=0.
\ee
where $u_2=-\frac{z_1}{1+R_3}$ and $v_2=\frac{z_1}{z_2}$.
Periodicity in the co-ordinate $t$ enforces  constraints on  the parameters of the solution. 
Using (\ref{feq3})  we can eliminate say the parameter $k_1$ in terms of $c_1, c_2, \omega$. 
We will see that it is crucial to impose periodicity in time to obtain the dispersion relation
for these strings.

The energy $\Delta$ and spin ${\cal S}$ of the string, which are the conserved
charges corresponding to time translations and shifts in $\phi$
are given by the following formulae respectively
\begin{eqnarray}
\label{energy}
{\Delta}&=&\sqrt{\l} E=\sqrt{\l} \int_0^{2\pi} \left[(1+r^2) c_1-b r^2 \phi'\right]\frac{d\s}{2\pi},\\
{\cal{ S}}&=&\sqrt{\l} S=\sqrt{\l} \int_0^{2\pi}\left[r^2 c_2 - b r^2 t'\right]\frac{d\s}{2\pi}.
\end{eqnarray}
An important point to note is that the integrands in the above expressions for the 
conserved charges are independent of the world sheet co-ordinate $\tau$. 
Therefore one can perform the integral for an arbitrary length in the 
$\sigma$ direction and still expect a conserved quantity. 
The angular momentum corresponding to rotations in $S^3$ is given by 
\begin{equation}\label{defj}
 J = \sqrt{\lambda} \omega .
\end{equation}
Using similar  change of variables and manipulations  as done to obtain
(\ref{eq1}) we can write the 
integral in the expression for the energy  given in (\ref{energy}) as 
\begin{equation}
E=  (c_1 + b k_2 ) + c_1 ( 1- b^2) \left(  R_3 + \frac{z_1}{2 \sqrt{A z_2} }  
{}_2F_1( \frac{1}{2}, \frac{3}{2} , 2, \frac{z_1}{z_2} ) \right).
\end{equation}
To further simplify the expression for the spin, it is convenient to 
relate the $S$ and the energy $E$.  
From the equations in (\ref{energy}) we obtain 
\be
\label{Disp1}
E-\frac{c_1}{c_2}S=c_1-b\int^{2\pi}_0(r^2 \phi'-r^2 t' \frac{c_1}{c_2})\frac{d\s}{2\pi} .
\ee
Using \ref{tp} and \ref{vir0} we can simplify the expression in the 
integrand. This results in the following
\be\label{niciden}
r^2 (\phi'-t'\frac{c_1}{c_2})=\frac{c_1}{c_2}\frac{b c_2 r^2-k_1}{1+r^2}=\frac{c_1}{c_2}t'.
\ee
Since the string is closed we have $t(\sigma+ 2\pi) = t(\sigma)$.  Therefore
the integral of  the above expression from $0$ to $2\pi$ vanishes. 
Thus the dispersion relation between energy and spin  takes the following form
\be\label{Disp2}
E=\frac{c_1}{c_2}S+c_1.
\ee
We emphasize the fact that the above dispersion relation is true only 
when one imposes the fact the string is closed in the $t$ direction. 
Closure in $\phi$ direction is not crucial for deriving the dispersion relation. 
Using the above relation and (\ref{Disp1}) we can write the following equation for the 
spin
\begin{equation} \label{sp}
 S = bk_1 + c_2 ( 1-b^2) \left(  R_3 + \frac{z_1}{2 \sqrt{A z_2} }  
{}_2F_1( \frac{1}{2}, \frac{3}{2} , 2, \frac{z_1}{z_2}) \right).
\end{equation}
Naively the dispersion  relation in (\ref{Disp2})  
does not involve the Neveu-Schwarz field $b$. But  as we will 
see subsequently we can use (\ref{eq1}) , (\ref{feq3})  and  (\ref{sp})  
to eliminate the independent parameters  $c_1, c_2, k_1$  
in favour of the spin $S$. 
We will derive this dispersion relation perturbatively  to order $b^2$.

\subsection{Perturbation theory in $b$} 

We have formally written the conditions for the  general solution of the equations of motion
of the spinning string in presence of the NS-NS field. In this section  we show that these conditions 
can indeed be satisfied by setting up a perturbative expansion in $b$. 
We will show that the crucial condition (\ref{feq3}) can be satisfied at the linear order in $b$. 
The condition (\ref{feq2}) which states that the closed string must be wound integer times
will  also be shown to be satisfied explicitly in the scaling limit 
in section \ref{scaling} at the linear order in $b$,

To proceed further we derive the corrections to the roots $R_1, R_2, R_3$ to order $b^2$ 
assuming the expansion (\ref{pexpand}).  These are given by 
\be\label{roots}
 R_1=-1+a_2 b^2 + O(b^3) ,\quad R_2=\frac{c_1^2-\w^2}{c_2^2-c_1^2}+a_1 b^2 + O(b^3) ,
 \quad R_3=a b^2 + O(b^3) , 
 \ee
 where
 \begin{eqnarray}\label{valroot}
 a_2&=&-\frac{(c_2+k_{1}^{(0)}  )^2}{c_2^2-\w^2},\\ \nonumber
a_1&=&- \frac{\w^2(c_1^2(k_{1}^{(0)}+ c_2)-c_2(c_2 k_{1}^{(0)}+ \w^2))^2}{c_2^2(c_2^2-c_1^2)(c_1^2-\w^2)(c_2^2-\w^2)} ,\\ \nonumber
a&=&\frac{c_1^2 (k_{1}^{(0)})^2}{c_2^2(c_1^2-\w^2)}.
  \end{eqnarray}
Recall that the turning points are at $r=\sqrt{R_3}$ and $r=\sqrt{R_2}$.
From the expression  in (\ref{Disp1})  and the identity (\ref{niciden}) 
we see that to obtain the dispersion relation 
to order $b^2$ it is sufficient to satisfy the closed string boundary conditions
to linear order in $b$ in the $t$ direction. This is because $t'$ occurs with a factor of $b$ in 
(\ref{Disp1}). Therefore satisfying the closed string boundary conditions to order $b$ will 
ensure that these terms start at order $b^3$ and therefore are of higher order in the 
dispersion relation.

Let us now examine the condition (\ref{feq3}) to order $O(b)$. 
From (\ref{tp}) we have 
\begin{eqnarray}
t' &=& bc_2 - \frac{b c_2 + k_1}{1+ r^2} , \\ \nonumber
&=& b c_2 - b \left( \frac{ c_2 + k_1^{(0)} }{1+ r^2}  \right) ,
\end{eqnarray}
where in the second line of the above equation we have kept terms to the linear order in $b$. 
Integrating  the world sheet 
co-ordinate $\sigma$ from $0$ to $2\pi$  the condition (\ref{feq3}) can be written 
\begin{equation}\label{tcondp}
2\pi   b c_2  - b ( c_2 + k_1^{(0)}) \int_0^{2\pi} \frac{d\sigma}{ 1+ r^2}  = 0  . 
\end{equation}
We can convert the integration over $\sigma$ to over $r$ using (\ref{vir2}). 
Performing similar change of variables as discussed earlier in the paper 
and working to the leading order in $b$ we obtain the condition
\be
\label{b11}
 c_2^{(0)}-\frac{(c_2^{(0)}+k_{1}^{(0)})}
 {\sqrt{(c_2^{(0)})^2-(c_1^{(0)})^2}}{}_2F_1(\frac{3}{2},\frac{1}{2},1,-\frac{(c_1^{(0)})^2 - \omega^2}
 {(c_2^{(0)})^2 - (c_1^{(0)})^2 } ) = 0. 
\ee 
where the superscript $(0)$ to indicate the zero order 
contributions of $c_1$ and $c_2$. We have integrated over $r$ between the turning points $\sqrt{R_3}$ and $\sqrt{R_2}$. 
Since there is an overall  factor of $b$ in the condition given in (\ref{tcondp})  these
turning points and all other terms are multiplying the equation are evaluated at the 
zeroth order in $b$. 
Finally to write the equation in (\ref{b11}) we have factored out the overall $b$. 
This equation  can be used to solve $k_1^{(0)}$ in terms of $c_1^{(0)}, c_2^{(0)}, \omega$. 

Let us now examine the leading behaviour of the  difference $\phi(2\pi) -\phi(0)$ given in 
(\ref{feq2}).   After a re-scaling of the variables by a change of variables  
the integral in (\ref{phint})  reduces to 
 \begin{eqnarray}
\label{phic2}
\int_0^{2\pi}\frac{d\s}{r^2}&&=\frac{4}{\sqrt{A R_3 } }\int_{1}^{\sqrt{\frac{R_2}{R_3} }}
\frac{dy}{y \sqrt{ ( R_3y^2 -R_1)(R_3 y^2 -R_2)(y^2-1)}}. 
\end{eqnarray}
Note that the upper limit of the integral in the limit $b\rightarrow 0$ tends to infinity since
$R_3 \sim O(b^2)$. 
To obtain the leading contribution of this integral  we can perform a taylor series 
expansion of the factor $( R_3y^2 -R_1)(R_3 y^2 -R_2)$ in $b$. 
The leading term is given by 
\begin{eqnarray} \label{manu}
k_2 \int_0^{2\pi}\frac{d\s}{r^2} &=&  \frac{4 k_2}{\sqrt{A R_3 R_1 R_2 }}
\left(   \int_{1}^{\infty }
\frac{dy}{y \sqrt{y^2 -1} }  + O(b^{1} )  \right),   \\ \nonumber
&=&     \left[   \left.  
-\tan^{-1}\left( \frac{1}{ \sqrt{y^2 -1}}\right)  \right|_{1}^{ \infty } + O(b^{1} )  \right],   \\ \nonumber
&=& 
  4( m + \frac{1}{2} ) \frac{\pi}{2}   +  4  \frac{k_2}{|k_2|} g( c_1^{(0)}, c_2^{(0)}, k_1^{(0)} )  + O(b^2) .  \nn
\end{eqnarray}
Here we have called the linear term in $b$ the function $g$ which depends on the zeroth order 
coefficients $ c_1^{(0)}, c_2^{(0)}, k_1^{(0)}$ \footnote{$ k_2^{(0)}$ can be determined 
using the Virasoro constraint.}, 
$m$ is any integer and in the last line we have used the relation  $A R_1 R_2 R_3 = k_2^2$. 
Thus the  difference in the end points of the string  in (\ref{feq2}) reduces to
\begin{equation}  \label{newclos}
\phi(2\pi) - \phi(0) = 
2\pi b c_1^{(0)}  - (2m + 1) \pi  - 4 \frac{k_2}{|k_2|} g( c_1^{(0)}, c_2^{(0)}, k_1^{(0)} ) + O ( b^2) .
\end{equation} 
Now we need to solve $k_1^0$ from equation (\ref{b11}) and then substitute 
in (\ref{newclos}) and check whether one obtains $2n\pi$. In general 
the difference
\begin{equation} \label{diffphi}
\delta \phi = 2\pi b c_1^{(0)}  -  b \frac{k_2^{(0)}}{|k_2^{(0)}|} g( c_1^{(0)}, c_2^{(0)}, k_1^{(0)} ).  
\end{equation}
will not vanish and therefore the string will not be closed for a single period.
 This situation is  similar to the 
giant magnon solution of \cite{Hofman:2006xt}.   We will show that we can 
construct a closed string solution after sufficiently large number of 
periods in $\sigma$.  That is we consider 
 \begin{equation}
 \phi( 2N\pi) - \phi(0) =  2m \pi + N \delta \phi,
 \end{equation}
 and we  we demand  
 $N\delta\phi = 2m'\pi$, $N, m'$ are integers.
 This implies that $\delta\phi$ is a rational multiple of $\pi$. 
  We will explicitly discuss this method of obtaining a closed string solution 
 in the scaling limit in section \ref{scaling} were we find the function $g$. 
 We will also see that $\delta\phi$ can be chosen to be a rational multiple of $\pi$. 
 So for the purposes of this paper we look for open string solutions in the $\phi$ direction for
 a single world sheet period, 
 but closed in the $t$ direction. We will  assume that a closed string in $\phi$ can be constructed.

The strategy to obtain   the dispersion relation is first solve $k_1^{(0)}$ in terms of 
$c_1^{(0)}, c_2^{(0)}, \omega$ using (\ref{b11}).  Then we can use (\ref{eq1}) to solve for say $c_2$ in 
terms of $c_1$ and $\omega$.  We use the equations  (\ref{sp}) and 
(\ref{defj}) 
to eliminate $c_1$ and $\omega$ in favour of the spin $S$ and angular momentum $J$.
Finally we substitute these values of $c_1, c_2, \omega$ in the relation for the 
energy in (\ref{Disp2}) to obtain the dispersion relation in terms of the spin and angular momentum. 
All these relations involve hypergeometric functions and therefore inverting them
is possible in certain limits.  
We will now restrict ourselves to three limits, the long  string, the scaling limit and the small string
 in which 
these functions simplify.

\subsection{Long string  limit }

Let us first consider the long string limit. 
This limit is obtained by pushing the length of the string  proportional to 
the difference in the turning points 
$z_1 = R_2  - R_3$ to infinity with $R_3$ held fixed. 
This is achieved by taking the parameters $c_2$  and $c_1$ to be almost equal. 
From now on we will restrict ourselves to the case in which $c_1, c_2, k_1, k_2, b$ all 
are positive. 
We will set $\omega =0$ to simplify our calculations. 
It is straight forward to repeat the analysis with $\omega\neq 0$. 
Under this limit, the hypergeometric functions simplify to 
\begin{eqnarray}\label{hyper}
{}_2F_1(\frac{1}{2},\frac{3}{2};2,\frac{z_1}{z_2})&\simeq &\sqrt{-\frac{z_2}{z_1}}\frac{4}{\pi}, \\ \nonumber
{}_2F_1(\frac{1}{2},\frac{1}{2};1,\frac{z_1}{z_2})&\simeq &\sqrt{-\frac{z_2}{z_1}}\frac{1}{\pi}\log({-\frac{z_1}{z_2}}). 
\end{eqnarray}
We first solve the the equation for $k_1^{(0)}$.  Using the asymptotic forms for the hypergeometric 
functions in (\ref{b11})   solve for $k_1^{(0)}$. This results in 
\begin{equation} \label{k10sol}
k_{1}^{(0)} = c_2^{(0)} \left( \frac{ \pi c_1^{(0)}}{2} -1  \right) . 
\end{equation}
Note that here the superscript $(0)$ to indicate the zero order 
contributions of $c_1, c_2$. 
Using the asymptotic forms
 of the hypergeometric functions and working to order $b^2$  the equation (\ref{eq1})
 reduces to 
 \begin{eqnarray}
\label{m21}
{b^2 \left(\pi  c_1^3 \left(( k_{1}^{(0)}) ^2-c_2^2\right)+2 c_1^2 (c_2+k_{1}^{(0)} )^2-
\pi  c_1 c_2^2 (k_{1}^{(0)})^2+2 c_2^2 ( k_{1}^{(0)})^2\right)} \\ \nonumber
= 2 (c_1 c_2)^2 \log\left( \frac{ c_1^2}{c_2^2 - c_1^2} \right)  - 2\pi c_1^3 c_2^2 . 
\end{eqnarray} 
Note that $k_1^{(0)}$ always occurs with terms suppressed by $b^2$, therefore 
we can substitute for it  from equation (\ref{k10sol}).
We can now solve $c_2$ in terms of $c_1$ to  $O(b^2)$. 
This gives 
\begin{eqnarray}\label{c2m2}
c_2 &=& \sqrt{e^{-\pi  c_1} c_1^2+c_1^2} +
\frac{b^2}{16 \sqrt{e^{-\pi c_1}+1}} \times \\ \nonumber
& & \left\{ c_1 e^{-2 \pi  c_1} (\pi  c_1-2) \left((\pi  c_1-2)^2+e^{\pi  c_1} (4-4 \pi  c_1)\right) \right\},
\end{eqnarray}
We then parametrize $c_1$ as 
\be\label{c1m2}
c_1=\frac{\log \left(\frac{1}{\n}\right)}{\pi }.
\ee
Substituting this parametrization of  $c_1$ into (\ref{sp}) and rewriting $c_2$ in terms of $c_1$ using (\ref{c2m2}) 
we obtain 
\begin{eqnarray}
S &=& \frac{2 \sqrt{\n+1}}{\pi  \n}
- \frac{b^2}{8 \pi  \n \sqrt{\n+1}} \times \\ \nonumber
& &\left\{ \log (\frac{1}{\n}) \left(4 (\n+1) (\n+6)-\log (\frac{1}{\n}) \left[\n^2 \log (\frac{1}{\n})
+10 \n+8\right]\right)-8 (\n+1) \right\}.
\end{eqnarray}
Here we have kept terms to $O(b^2)$. 
We can now solve for $\nu$ in terms of $S$ to $O(b^2)$, this leads to 
\begin{eqnarray} \label{nm2}
\n &=&\frac{2}{\pi S} \sqrt{ 1 + \frac{1}{\pi^2 S^2} } + \frac{2}{\pi^2 S^2}  
 + \frac{b^2}{2 \pi ^{5/2} S^{\frac{5}{2}}   \sqrt{{2}+\pi S }} \times \\ \nonumber
& & \left\{2 \pi  S (\pi  S+2)+\log ^3\left(\frac{\pi  S}{2}\right)+
\pi  S (2 \pi  S+5) \log ^2\left(\frac{\pi  S}{2}\right)  \right. 
\\ \nonumber & & \left.  -2 (\pi  S+2) (3 \pi  S+1) \log \left(\frac{\pi  S}{2}\right) 
\right\}.
\end{eqnarray}
Hence using (\ref{nm2}), (\ref{c2m2}) and (\ref{c1m2}) we can write down $c_1$ and $c_2$ in terms of $S$.
Substituting these in the expression for the energy given in (\ref{Disp2}) and working to the $O(b^2)$ we obtain 
\begin{eqnarray}\label{dispm2}
E&=&\frac{c_1}{c_2}S+c_1\\ \nonumber
&=&S + \frac{1}{\pi} \log S 
-\frac{b^2}{2\pi}  \log^2 S   + O( (\log S)^0, b^2 \log S).
\end{eqnarray}
In the terms of the above equation we have kept the leading order in the spin $S$ at each order in the $b^2$ expansion.
For the coefficient of the $O(b^0)$ term we have neglected terms which are of $O(\log S)^0$. 
In the coefficient of $O(b^2)$ term we have neglected  terms which are of $O(b^2 \log S)$. 
We can rewrite this dispersion relation in terms of the physical charges as 
\begin{equation}
\Delta = {\cal S}  + \frac{\sqrt{\lambda}}{\pi} \log {\cal S} - 
\frac{b^2 \sqrt{\lambda}}{2\pi} \log^2 {\cal S} . 
\end{equation}
Therefore the dispersion relation of the large spinning string is corrected at $O(b^2)$. The leading correction 
at this order is given by $ - \frac{b^2\sqrt{\lambda} }{2\pi}  \log^2 {\cal S }$.  
Note that it is clear from our analysis that we have first performed a perturbation in $b$ and then 
at each order extracted out the leading behaviour in the spin ${\cal S}$. 
We will  arrive at the above dispersion relation in the scaling limit of the long string solution
in the next section.

\subsection{Scaling limit of the long string} \label{scaling}

There is a further limit of the long string solution in which the solution simplifies   \cite{Kruczenski:2007cy}. 
This limit is known as the scaling limit.  In this limit it is possible to write down the functional dependence of the 
the radial co-ordinate $r$ on the world sheet $\sigma$ in terms of a simple function rather than 
hypergeometric function or elliptic functions. 
The  scaling limit of the long string solution in the absence of the NS-NS B-field  has been mapped by 
$SO(2,2)$ transformations and world sheet  re-parametrizations
to the minimal surface corresponding to the cusp Wilson line \cite{Kruczenski:2007cy}. 
One of the goals of this paper is to 
obtain this mapping 
 in the presence of the NS-NS flux, 
With this motivation  we will study this scaling limit.

\subsubsection*{$b=0$}

Let us first review the scaling limit of \cite{Kruczenski:2007cy} with $b=0$ in our language which 
will enable the generalization to the situation with the NS-flux. 
From the equations (\ref{c2m2})  we see that,  the constants $c_1$ and $c_2$ are related by 
\begin{equation}
 c_2 = c_1  + O( c_1 e^{-c_1}) .
\end{equation}
Now in the large $S$ limit,   (\ref{c1m2}) and (\ref{nm2}) implies that $c_1$ is large  so 
we can ignore the exponentially 
suppressed term. 
Therefore we can look for a solution with $c_1 =c_2$ to begin with, this solution is the
scaling limit of the long string.  
From (\ref{tp}) and $b=0$ we have
\begin{equation}
 t= c_1 \tau, \qquad \tilde \phi = c_1 \tau. 
\end{equation}
The differential equation for $r$  given in (\ref{vir1}) reduces to \footnote{We have set $\omega =0$.}  
\begin{equation} \label{scali}
 r'^2 = c_1^2 ( r^2 +1) .
\end{equation}
To ensure that the solution is a closed folded string, the solution is allowed to grow from 
$r=0$ to $r_{\rm max}$ from $\sigma =0$ to $\sigma = \pi/2$, then $r$ decreases back to 
zero as $\sigma$ goes from $\pi/2$  to $\pi$. 
The same motion repeats for the interval $\pi<\sigma \leq 2\pi$. 
Integrating the equation (\ref{scali}) we obtain 
\begin{equation}
 r = \sinh( c_1 \sigma), \qquad 0\leq \sigma \leq \frac{\pi}{2} .
\end{equation}
The energy and spin of the solution is given by 
\begin{equation}
 E = \frac{4c_1}{2\pi} \int_0^{r_{\rm max} } \sqrt{ 1+ r^2} dr , \qquad
 S =  \frac{4c_1}{2\pi}   \int_0^{r_{\rm max} }  \frac{r^2}{ \sqrt{1 + r^2} } dr.
\end{equation}
Here we have used the fact that $S$ receives contribution from the $4$ segments of the 
folded string and used (\ref{scali}) to write the integral in terms of the radial co-ordinate.
Integrating the equation for $S$ and using the fact that $c_1$ is large we obtain
\begin{equation}
S \sim \frac{1}{2\pi} e^{c_1\pi}.
\end{equation}
We can now substitute for $c_1$ in the dispersion relation (\ref{Disp2}) to obtain 
\begin{equation}
E - S = \frac{1}{\pi} \log S
\end{equation}
Re-writing this in terms of physical charges we obtian
\begin{equation} \label{longdisp}
\Delta - {\cal S} = \frac{\sqrt{\lambda}}{\pi} \log {\cal S}.
\end{equation}
Note that  in the limit $c_1 = c_2$  we have obtained the dispersion 
relation of the long string.  Further more  there exists a simple and explicit expression
for the solution for the solution $r(\sigma)$. 

\subsubsection*{ $b\neq 0$}

Let us now generalize the scaling limit  in the presence of the NS-flux. 
As before we choose  the following  ansatz for the solution 
\be
t=c_1\t+t(\s),\quad \tilde \phi=c_2\t+\phi(\s),\quad r=r(\s), 
\ee
This ansatz is the same as the one considered in the previous section but  with $c_1=c_2$. 
The Virasoro constraints (\ref{vir0}) then reduces to 
\be\label{viros}
k_1=k_2
\ee
From (\ref{tp}) we see that the  
conservation laws for $t$ and $\phi$ is given by 
\be\label{tpsc}
t'=\frac{b c_1 r^2-k_1}{1+r^2},\quad \phi'=\frac{b c_1r^2-k_1 }{r^2}. 
\ee
As we have discussed in the previous section   we look for a solution  in which $k_1$ 
admits a power series in $b$ as given in (\ref{pexpand}). 
Finally  the Virasoro equation (\ref{vir1}) takes the following form
\be\label{vir1sc}
r'^2r^2= \tilde B r^4+\tilde Cr^2+\tilde D ,
\ee
where
\begin{eqnarray}
\label{Bcd}
\tilde B= c_1^2-c_1^2 b^2 , \qquad
\tilde C=c_1^2+2bc_1k_1 ,\qquad
\tilde D=-k_1^2.
\end{eqnarray}
The equation for $r'$ simplifies to a quadratic polynomial in $r^2$. 
Let the two roots of the polynomial be $\tilde R_1$ and $\tilde R_2$, then  
we have
\begin{equation} \label{scalrp}
 r' = \frac{1}{r} \sqrt{ \tilde B ( r^2 - \tilde R_1) ( r^2 - \tilde R_2) },
\end{equation}
where to $O(b^2)$ the roots are given by 
\begin{equation}\label{rootsca}
 \tilde R_1 = \frac{( k_1^{(0)})^2}{(c_1^{(0)}) ^2} b^2 , \qquad 
 \tilde R_2 = -1  - \frac{ (c_1^{(0)} + k_1^{(0)})^2}{(c_1^{(0)})^2 } b^2.
\end{equation}
The turning points to integrate the equation for $r'$ is  root $R_1$ and  the point $r_{\rm max}$ 
which  is reached at $\sigma = \pi/2$. 
The solution for $0<\sigma <\pi/2$ is given by 
\begin{eqnarray}\label{r2sc}
r^2=\frac{e^{-2 \sqrt{\tilde B} \sigma }}{4 \a \tilde B} 
\left\{\left(\a e^{2 \sqrt{\tilde B} \sigma} - \tilde C  \right)^2-4 \tilde B \tilde D\right\}, \\ \nonumber
\a=2 \sqrt{\tilde B} \sqrt{\tilde B \tilde R_1^2+\tilde C \tilde R_1+\tilde D}+2 \tilde B \tilde R_1+\tilde C.
\end{eqnarray}
It can be seen that on setting $b=0$, this solution reduces to $r = \sinh c_1 \sigma$. 

 Let us now study the closed string boundary conditions $t(\sigma ) = t(\sigma + 2\pi)$. 
 As discussed in the previous section to obtain dispersion relation to $O(b^2)$ it is sufficient to 
 study this constraint to the linear order in $b$. 
 Integrating the equation for $t'$ in (\ref{tpsc}) and after imposing closed string boundary conditions we obtain
 \begin{equation}\label{scaltcl}
  \int_0^{2\pi} t' d\sigma = 4  \int_{\sqrt{\tilde R_1}}^{ r_{\rm max}}  \left (  b c_1 - \frac{bc_1 + k_1}{ 1+ r^2}   \right) 
  \frac{dr}{r'}   =0.
 \end{equation}
 To the linear order in $b$, the above equation reduces to 
 \begin {equation}
  b \int_0^{r_{\rm max}} \left( c_1^{(0)} -  \frac{c_2^{(0)} + k_1^{(0)}} {1+r^2}\right)  \frac{dr}{\sqrt{1 + r^2}} =0.
 \end {equation}
 Note that we have substituted the zeroth order values for the limits and for $r'$. 
 This equation can be easily integrated
 by change of variables to the $\sigma$ coordinate. This leads to the following equation for $k_1^{(0)}$
 \be
 b c_1^{(0)} \frac{\pi}{2}-\frac{b c_1^{(0)}+b k_{1}^{(0)}}{c_1^{(0)}}\tanh\left(c_1^{(0)}\frac{\p}{2}\right)=0. 
 \ee
 Since we are working in the large string limit we can approximate $\tanh(c_1^{(0)}\frac{\p}{2})\sim 1$.
 Therefore we obtain
 \be \label{k10sols}
 k_{1}^{(0)}=\frac{( c_1^{(0)})^2 \pi}{2}-c_1^{(0)}. 
 \ee
 Note that this is identical to the value obtained in the previous section in equation  (\ref{k10sol}) without taking the 
 scaling limit.     Thus we obtain $k_1^{(0)} = k_2^{(0)} >0$ since $c_2^{(0)} = c_1^{(0)}$ is large. 
 Let us now integrate the equation for $\phi'$, the solution is given by 
 \begin{eqnarray}
 \tilde\phi&=&  c_1 \tau + \int_0^\sigma \phi' d\sigma, \\ \nonumber
 &=& c_1 \tau + \int_{\sqrt{R_1}}^{r(\s)}\frac{\phi'}{r'}dr. 
 \end{eqnarray}
 Substituting the expression for $r', \phi'$ and integrating we obtain 
 \begin{equation}
 \tilde \phi = c_1\tau + c_1 b \sigma  -\frac{k_2}{|k_2|} \tan^{-1} 
 \left(\sqrt{|\frac{R_2}{R_1}|}\sqrt{\frac{r^2(\s)-R_1}{r^2(\s)-R_2}}\right). 
 \end{equation}
 To obtain this solution we have used the relation
 $k_2^2  = -\tilde B R_1 R_2 $ which follows from the definition of the roots of the 
 quadratic polynomial in (\ref{vir1sc})  and the Virasoro constraint (\ref{viros}). 
 To the leading order in $b$, the solution is given by 
 \begin{equation} \label{solphis}
 \phi = c_1^{(0)} \tau + c_1^{(0)} b \sigma  - \tan^{-1}  
 \left( \frac{c_1^{(0)} }{ b k_1^{(0)}} \tanh ( c_1^{(0)} \sigma) \right)  . 
 \end{equation}
 Therefore the difference in the end points in the $\phi$ direction for a single period is given by 
 \begin{equation}
 \phi( 2\pi) - \phi(0) = 2\pi c_1^{(0)} b - ( 2m + 1) \frac{\pi}{2} +4 b   \frac{k_1^{(0)} }{ c_1^{(0)} }. 
 \end{equation}
 Here we have used the fact that the integral over $r$ can be broken up to 4 integrals 
 from $0$ to $r_{\rm max}$ and $r_{\rm max}  = \sinh ( c_1^{(0)} \frac{\pi}{2} ) \sim   \cosh ( c_1^{(0} \frac{\pi}{2} )$ since $c_1^{(0)}$ is large. 
 From the definition of $\delta \phi$ in (\ref{diffphi}) we obtain 
 \begin{eqnarray}
 \delta \phi & = & 2\pi c_1^{(0)} b  + 4 b   \frac{k_1^{(0)} }{ c_1^{(0)} } , \\ \nonumber
 &=& ( 4\pi c_1^{(0)}  - 4 ) b.
 \end{eqnarray}
 In the second line of the above equation we have used (\ref{k10sols}). 
 Therefore in general the string does not close in the $\phi$ direction analogous to the 
 giant magnon solution of \cite{Hofman:2006xt}.  But if one considers the difference
 after $N $ periods with $N$ large enough in the $\sigma$ direction we can can
 make the difference and integer multiple of $2\pi$. That is we can have
 \begin{equation}\label{cccond}
 N\delta\phi = N( 4\pi c_1^{(0)} -4) b = 2m' \pi, 
 \end{equation}
  where $m'$ is an integer. 
  We see that we can ensure this mild quantization condition on $\delta \phi$ by either 
  choosing $b$ or $c_1^{(0)}$ to satisfy the above condition. 
 Note that it does not put any restriction on $c_1^{(0)}$ or $b$, that is 
 we can still have $c_1^{(0)} \rightarrow\infty$, $b<<1$ and 
 $ c_1^{(0)} b < <1$. 
 This is seen as follows, since we are working in the large $c_1^{(0)}$ limit the condition 
 in (\ref{cccond}) reduces to $2N c_1^{(0)}b = m $. By choosing $N$ sufficiently large we can 
 ensure that the condition $ c_1^{(0)} b < <1$ necessary for perturbation theory is satisfied
 \footnote{We need  $ c_1^{(0)} b < <1$ so that $b \log {\cal S} <<1$,  this ensures that
 the perturbative expansion in the dispersion relation makes sense.}.
 This implies we can make the string closed after a sufficiently large period in the 
 $\sigma$ direction.

By using the expression for $t'$ given in (\ref{tpsc}) 
the equation for the spin $S$  in (\ref{energy}) can be written as 
\be
S=\int_0 ^{2\pi}(r^2 c_1(1-b^2)+b k_1)\frac{d\s}{2\pi}.
\ee
Here we have using the closed string boundary condition on the co-ordinate $t$ \footnote{We have satisfied 
this to the linear order in $b$ in (\ref{scaltcl})}. 
Substituting the explicit solution for $r(\sigma)$ given in (\ref{r2sc}) and performing the integration 
in $\sigma$ we obtain the following expression for the spin to $O(b^2)$. 
\begin{eqnarray} \label{spinscexp}
2\pi S&=&\ (\sinh (\pi  c_1)-\pi  c_1)+\\ \nn
&&\frac{b^2}{2 c_1^{2}} 
\left(-\pi  c_1^3 \cosh (\pi  c_1)+
c_1^2 \sinh (\pi  c_1)+4 k_{10}^2 \sinh (\pi  c_1)+4 c_1 k_{1}^{(0)} \sinh (\pi  c_1)\right) . 
\end{eqnarray}
We can now solve for $c_1$ perturbatively  in $b$  by assuming an expansion of the form
$c_1 = c_{1}^{(0)}+b^2 c_{1}^{(1)} $. 
Substituting this expansion in (\ref{spinscexp}) and matching orders to 
$b^2$ we obtain
\begin{equation} \label{c11sc}
 c_1^{(0)} = \frac{\log S}{\pi}, \qquad c_1^{(1)} = -\pi \frac{(c_{1}^{(0)})^2}{2} = -\frac{1}{2\pi} \log^2 S.
\end{equation}
We have taken the large $c_1^{(0)}$ limit when we solve for each order in $b$
\footnote{Note here again we have first performed the perturbation 
in $b$ and then at each order in $b$ retained the contribution from the spin.}. We have also used the 
value of $k_1^{(0)}$ obtained in (\ref{k10sols}). 
Substituting this value of $c_1$ to order $b^2$ into the dispersion relation , 
we obtain
\begin{equation}
 E-S=c_{1}^{(0)}+b^2 c_{1}^{(1)}=\frac{\log S}{\pi}-b^2 \frac{\log^2 S}{2\pi}. 
\end{equation}
Note that after expressing this dispersion relation 
in terms of physical charges agrees  with that obtained in (\ref{longdisp}) for the spinning string 
without the scaling limit.

The scaling limit enables one to find a closed form expression for the solution 
as a function of $\sigma$ to $O(b^2)$. 
We can write down the solutions explicitly to the order of $b^2$ now. Using (\ref{r2sc}) we write down the solution in the global coordinate $r$ and $\rho=\sinh^{-1} r$ in order of $b$.
\begin{eqnarray}
r&=&\sinh (c_{1}^{(0)} \sigma )+\frac{1}{8} b^2 \big\{4 \sinh (c_{1}^{(0)} \sigma )-
4 c_{1}^{(0)} (\pi  c_{1}^{(0)}+1) \sigma  \cosh (c_{1}^{(0)} \sigma )+\\ \nn
&&(\pi  c_{1}^{(0)}-2) (\pi  c_{1}^{(0)} \cosh (2 c_{1}^{(0)} \sigma )-2) 
\text{csch}(c_{1}^{(0)} \sigma )\big\}, \\ \nn
\rho&=&c_{1}^{(0)}\s+\frac{1}{8} b^2 \left\{c_{1}^{(0)} \left(\pi ^2 c_{1}^{(0)} \tanh (c_{1}^{(0)} \sigma )
-4 (\pi  c_{1}^{(0)} \sigma +\sigma )\right)+(\pi  c_{1}^{(0)}-2)^2 \coth (c_{1}^{(0)} \sigma )\right\}.
\end{eqnarray}
In writing down the expansions in $b$ we have assumed $\sigma$ is sufficiently away from 
$0$. If $\sigma$ is close to zero one needs to use the expression in (\ref{r2sc}) directly. 
Now plugging in the explicit value of $r$ in (\ref{tpsc}) and using the vlue of $c_{1}^{(1)}$ from 
\ref{c11sc} in the expansion of $c_1$, we obtain the expression of $t$ and $\phi$ in terms of $\t$ and $\s$.
\begin{eqnarray}
\tilde t=c_{1}^{(0)}\t-\frac{b^2 (c_{1}^{(0)})^2\pi}{2}\t+\frac{1}{2} b c_{1}^{(0)} 
\left\{2 \sigma -\pi  \tanh (c_{1}^{(0)} \sigma )\right\} . 
\end{eqnarray}
The solution for $\phi$ can be obtained as an expansion in $b$ from (\ref{solphis}). 
This results in 
\begin{equation}
\tilde\phi = c_1^{(0)} \tau -\frac{b^2 (c_1^{(0)})^2\pi}{2}\t+ c_1^{(0)} b \sigma  + \frac{b}{2} ( \pi c_1^{(0)} - 2) \coth(  c_1^{(0)} \sigma) -b(\pi c_1^{(0)}-1).
\end{equation}
Note that again this expansion is valid for $\sigma$ sufficiently away from the origin, 
if $\sigma$ is close to the origin we can use the solution in (\ref{solphis}). 
We have chosen an integration constant in $\phi$ such that $\phi(\tau =0, \frac{\pi}{2})$ 
vanishes for $c_1^{(0)}$ large. This done for convenience in later manipulations.

Let us now re-write the   solution is obtained  after scaling the worldsheet coordinates 
by redefining  $(\s',\t')=(c_1^{(0)} \s,c_{1}^{(0)} \t)$. This rescaling of $\s$ with $c_1^{(0)}\rightarrow \infty$ effectively de-compactifies $\s$. After the scaling the solutions look like the following
\begin{eqnarray}\label{scalingsol}
\r&=&\s+\frac{1}{8} b^2 \left\{\pi ^2 (c_1^{(0)})^2 \tanh (\sigma )-4 (\pi  c_1^{(0)}+1) \sigma +(\pi  c_1^{(0)}-2)^2 \coth (\sigma )\right\}, \\ \nonumber
\tilde t&=&\t-\frac{1}{2} \pi  b^2 c_1^{(0)} \tau + b\s -\frac{1}{2} \pi  b c_1^{(0)} \tanh (\sigma ), \\ \nonumber
\tilde \phi&=&\t-\frac{1}{2} \pi  b^2 c_1^{(0)} \tau +b\s+\frac{1}{2} b (\pi  c_1^{(0)}-2) \coth (\sigma )-b(\pi c_1^{(0)}-1). 
\end{eqnarray}
where we have dropped the primes from $\t$ and $\s$. One can check that this satisfies all the equations of motions and the Virasoro constraints of the  sigma model.
We will comment on the choice of integration constants in the solution
given in (\ref{scalingsol}) later.

\subsection{Small string limit}

We now consider the opposite limit, that is the limit in which the extent of the string in the 
$r$ direction is small. 
To obtain the limit we first set $k_1 = k_2=0$.  The second Virasoro constraint 
given in (\ref{vir0}) is 
automatically satisfied. 
Using the Virasoro constraint corresponding to the vanishing of 
the world sheet energy for this situation leads to the following 
equation for the radial co-ordinate $r$
\be
\label{Virsim}
r'^2 =\hat A r^4+\hat B r^2+C = \hat A( r^2 -\hat R_1) ( r^2-\hat R_2), 
\ee
where
\be
 \hat A=c_1^2+c_2^2 b^2-c_2^2-c_1^2 b^2 ,\quad
\hat B=2 c_1^2-c_2^2-c_1^2 b^2-\w^2,\quad
\hat C=c_1^2-\w^2 . 
\ee
The situation now is analogous to the case of $b=0$ in which there were only $2$ roots. 
The  roots of the quadratic equation in (\ref{Virsim}) to order $b^2$ are given by 
\be\label{smalro}
\hat R_1=-\frac{b^2 c_2^2}{c_2^2-\w^2}-1,\quad \hat R_2=\frac{b^2 \w^2 \left(c_1^2-\w^2\right)}{\left(c_1^2-c_2^2\right) \left(c_2^2-\w^2\right)}+\frac{\w^2-c_1^2}{c_1^2-c_2^2}. 
\ee
The solution for $r(\sigma)$ now begins at the origin reaches $\sqrt{R_2}$ at $\sigma = \pi/2$ and 
then turns back and returns to the origin in the next quarter period. The same behaviour is 
repeated in the interval $\pi<\sigma\leq 2\pi$. 
Therefore integrating the equation for $r$ leads to 
\be
2\pi =\int^{2\pi}_0d\s=4\int^{\sqrt{\hat R_2}}_0\frac{dr}{\sqrt{\hat A (r^2-\hat R_1)(r^2-\hat R_2)}}. 
\ee
After similar manipulations the integral can be performed in terms of the hypergeometric 
function and it leads to 
\be\label{rclosed}
\sqrt{\hat R_1 \hat A}={}_2 F_1 \left(\frac{1}{2},\frac{1}{2},1,\frac{\hat R_2}{\hat R_1}\right). 
\ee
The small string approximation is essentially the fact that the maximum extent of the 
string $\hat R_2\rightarrow 0$. In this limit, the hypergeometric function just reduces to 
$1$. Therefore we obtain the relation 
\begin{equation}\label{ra}
\hat R_1 \hat A =1. 
\end{equation}
Substituting the expressions for $\hat R_1,\hat  A$ from (\ref{smalro}) leads  to the 
constraint 
\be
\label{rclosedsim}
 c_2^2 - c_1^2 -1 + b^2 \omega^2\left(  \frac{ c_1^2 -c_2^2 }{\w^2-c_2^2}\right) =0.
\ee
Let us now examine if the closure condition for $t$ is satisfied. Integrating the equation 
for $t'$ in (\ref{tp}) with $k_1=0$ leads to 
\begin{equation}
t(2\pi) - t(0) = 2\pi b c_2 - bc_2 \int_0^{2\pi} \frac{d\sigma}{1+r^2} . 
\end{equation}
Again performing the same manipulations  in the integrand and examining the condition 
to the leading order gives rise to the equation
\be
 c_2^{(0)}-\frac{c_2^{(0)}}
 {\sqrt{(c_2^{(0)})^2-(c_1^{(0)})^2}}{}_2F_1(\frac{3}{2},\frac{1}{2},1, \frac{\hat{R_2}}{\hat{R_1}} ) = 0. 
\ee 
Approximating the hypergeometric function by unity since the $R_2\rightarrow 0$ and using 
the fact $(c_2^{(0)})^2-(c_1^{(0)})^2 =1$ from (\ref{rclosedsim}), we see that the condition for 
periodicity in $t$ is satisfied. 
The equation of motion for $\phi$ is just $\phi' = b c_1$, here again the 
string does not close, but as discussed earlier one can consider several periods and 
ensure the string closes in the $\phi$ direction. 

Let us now obtain the dispersion relation for the string. For this we need the 
expression for the spin which is given by 
\begin{eqnarray}\label{spin1s}
S& =& \int_{0}^{2\pi}\left[c_2 (1-b^2) r^2 \right] \frac{d\s}{2\pi},   \\ \nonumber 
&=&\frac{\hat{R_2} c_2 (1-b^2)}{2 \sqrt{A \hat{R_1}}} {}_2F_1\left(\frac{1}{2},\frac{3}{2},2,\frac{\hat{R_2}}{\hat{R_1}}\right), 
\\ \nonumber
&=& \frac{\hat{R_2}}{2}  c_2( 1-b^2). 
\end{eqnarray}
To obtain the first line we have used the closed string boundary condition on $t$. 
In the last line of the above equation we have approximated the hypergeometric 
equation by unity and also used (\ref{ra}). 
We now have all the ingredients to obtain the dispersion relation. 
We first parametrize $c_2$ by 
\begin{equation}
c_2^2 = 1 + \omega^2 + x. 
\end{equation}
From (\ref{rclosedsim}) we obtain $c_1$ as 
\be
c_1=\frac{b^2 \w^2}{2 (x+1) \sqrt{\w^2+x}}+\sqrt{\w^2+x}
\ee
We now examine the situation in which $x<<1$ studied earlier for the 
situation with $b=0$ in \cite{Frolov:2002av}. We plug these values of $c_1$ and $c_2$ in the equation for spin (\ref{spin1s}) and expand the expression to the linear order of $x$. 
\be
S=\frac{1}{2} b^2 \sqrt{\w^2+1} \w^2+\frac{x \left(-2 b^2 \w^4-3 b^2 \w^2-2 b^2+2 \w^2+2\right)}{4 \sqrt{\w^2+1}}
\ee
We then solve for $x$ in terms of $S$ and $\w$. To the order of $b^2$ it is
\be
x=\frac{2 S}{\sqrt{\w^2+1}}+\frac{b^2 \left(2 S \w^4+3 S \w^2+2 S-\sqrt{\w^2+1} w^2-\sqrt{\w^2+1} \w^4\right)}{\left(\w^2+1\right)^{3/2}}
\ee
So now we can write down $c_1$ and $c_2$ in terms of $S$ and $\w$ and now we can write down the dispersion relation.
\begin{eqnarray}
c_1&=&\sqrt{\frac{2 S}{\sqrt{\w^2+1}}+\w^2}+\\\nn
&& \frac{b^2 S \sqrt{\frac{2 S}{\sqrt{\w^2+1}}+\w^2} }{2 \left(\w^2+1\right) \left(2 S+\sqrt{\w^2+1}\right) \left(4 S^2 \sqrt{\w^2+1}+2 S \left(\w^2+1\right)^2+\w^2 \left(\w^2+1\right)^{3/2}\right)}\times\\ \nonumber
&&\left(4 S^2 \sqrt{\w^2+1} \left(2 \w^4+3 \w^2+2\right)+4 S \left(\w^6+3 \w^4+4 \w^2+2\right)+\sqrt{\w^2+1} \left(\w^4+3 \w^2+2\right)\right) \\ \nonumber
c_2&=&\sqrt{\frac{2 S}{\sqrt{\w^2+1}}+\w^2+1}+\frac{b^2 \left(S \left(2 \w^4+3 \w^2+2\right)-\w^2 \left(\w^2+1\right)^{3/2}\right)}{2 \left(\w^2+1\right)^{3/2} \sqrt{\frac{2 S}{\sqrt{\w^2+1}}+\w^2+1}} 
\end{eqnarray}
We then find the dispersion relation to the order of $b^2$ in the limit $\w>>1$
\begin{eqnarray}
E&=&\frac{c_1}{c_2}S +c_1 \\ \nonumber
&=&\w+S+\frac{S}{2\w^2}+\frac{b^2 S}{2}-\frac{b^2 S }{4 \w^2}+O(S^2)+O\left(\frac{1}{\w^3}\right)
\end{eqnarray}
 To obtain this we have neglected higher order terms in the spin $S$.
 This is  consistent in the small string limit since in this approximation 
 we have to neglect higher powers of $x$. 
 Let us now recast this dispersion relation in the conventional form by re-defining the 
  spin to $O(b^2)$ as
  \begin{equation}
  \tilde S = \left( 1 + \frac{b^2}{2}\right)   S 
\end{equation}
Then it terms of this rescaled spin\footnote{Note that this is just 
a redefinition of what is called the spin. }, we obtain the dispersion relation 
\begin{equation}
E = \omega + \tilde S + \frac{\tilde S}{2\omega^2} - b^2 \frac{\tilde S}{2\omega^2} .
\end{equation}
We can now write the dispersion relation in 
terms of macroscopic charges by re-instating $\sqrt{\lambda}$. 
This results in  
\begin{equation} \label{smafin}
\Delta - J = {\cal S}  + \frac{\lambda}{2J^2} ( 1- b^2) {\cal S}. 
\end{equation} 

The dispersion relation for the small string can be compared to that of the plane wave spectrum 
in presence of the mixed 3-form fluxes. This was derived in  \cite{Berenstein:2002jq}, the dispersion relation 
is given by \footnote{See equation (C.3) of \cite{Berenstein:2002jq}. }
\begin{equation}\label{bmn}
\Delta - J = \sum_n N_n
 \left(  1 + 2 \frac{\sqrt{\lambda} n}{J} \sin \beta   + \frac{ \lambda n^2}{J}  \right)^{\frac{1}{2}}. 
 \end{equation}
 We have defined $\beta = \frac{\pi}{2} - \alpha$,  
 where $\sin\beta$ is the coefficient of the Neveu-Schwarz flux. 
 We have also identified $\frac{\sqrt\lambda}{J} = \frac{1}{\alpha' p^+}$ and have chosen 
 $\mu =1$. 
 Let us now expand the square root in (\ref{bmn}) and also use the 
 level matching condition $\sum \limits_{n=-\infty}^{\infty}n N_n=0$. We then obtain 
 \begin{equation}\label{pwave}
 \Delta -J = \sum_n N_n +\left(\sum_n n^2 N_n\right) \frac{\lambda}{2J^2}  ( 1 - \sin^2 \beta)  + O(1/J^3) .
 \end{equation}
  Following \cite{Frolov:2002av} we identify  the charge ${\cal S}$ with the excited state
  having 
  quantum numbers $n=1, N_1 = \frac{\cal S}{2}, n=-1,  N_{-1} = \frac{\cal S}{2} $. We then see that the 
  plane wave dispersion relation (\ref{pwave})   precisely matches with that
  obtained from the small string in (\ref{smafin}) for Neveu-Schwarz flux 
  $\sin\beta = b$.

\section{Minimal surfaces with mixed form fields}

In $4$ dimensional conformal field theories the anomalous dimensions of 
high spin twist two operators is related to the logarithmic divergence 
of the expectation value of the Wilson loop which has a cusp in its contour. 
This relationship can be established entirely in the field theory. 
In the bulk it was shown in \cite{Kruczenski:2007cy} that the classical solution of spinning 
strings  in the scaling limit 
is related by conformal transformations and re-parametrization to the
minimal surface corresponding to the cusped Wilson loop. 
The $2$ dimensional conformal field theories dual to the 
$AdS_3\times S^3$ backgrounds are not as well understood as those in 
$4$ dimensions.  Therefore it is interesting to ask the question if the 
anomalous dimensions of high spin operators determines the cusp anomaly 
for these $2$ dimensional theories.  If this question is posed in the bulk  
$AdS_3$  background supported
 with purely RR 3-form flux then the same conformal transformations
and re-parametrizations 
found in \cite{Kruczenski:2007cy} is sufficient to relate the two classical solutions. 
This is because the spinning string and minimal surface can be embedded in 
$AdS_3$.   

It is less clear how to relate the  spinning string  in the $AdS_3$ background 
supported with NS-NS 3-form flux to minimal surfaces corresponding to cusped Wilson 
loops. 
   To establish this relation  in section \ref{wilsonloop} we first study minimal 
surfaces which end on a light like cusp. We show that the equation of motion for the 
minimal surface can be solved exactly to all orders in the NS-field for the special 
case called the `uniform'  minimal surface. 
 In general the equation of motion admits a solution in terms of a  perturbative expansion in $b$.
We then evaluate the area of the minimal surface and show that the coefficient of logarithmic 
divergence of the area  proportional to $b^2$ is precisely the coefficient 
of the $b^2 \log^2 S$ in the anomalous dimensions of the spinning string solution. 
In section \ref{relatewilspin} starting from the scaling limit of the spinning string  solution 
given in (\ref{scalingsol}) to $O(b^2)$ we perform $SO(2,2)$ transformations and re-parametrizations
to relate the solution to the minimal surface corresponding to the cusped Wilson loop. 
This minimal surface is a one parameter generalization of the `uniform' minimal surface depending
of the parameter $c_1^{(0)}$ of the spinning string solution in (\ref{scalingsol}).  
We will see  that  the logarithmic divergence of the area of the 
minimal surface is however universal and independent of this parameter and 
precisely the coefficient of $b^2 \log^2 S$  of the spinning string solution.

\subsection{Cusp anomalous dimensions  from gravity}\label{wilsonloop}

We look for minimal surfaces which end on a light like cusp in the Poincar\'{e} co-ordinates
of $AdS_3$ given by the following metric
\begin{equation}\label{poinmet}
ds^2 = \frac{1}{z^2} ( - dx_0^2  + dx_1^2 + dz^2) 
\end{equation}
The NS-NS 2-form  in this co-ordinate system is given by 
\begin{equation}
B_{x_0 x_1} =  - \frac{b}{z^2}
\end{equation}
It can be easily verified that this background is a solution to the bulk equations of motion. 
The minimal surface is a solution to the equations of the Nambu-Goto action given by 
\begin{equation}\label{ngact}
S_{NG} = \frac{\sqrt{\lambda}}{2\pi} \int d^2 \sigma 
\left( \sqrt{{\rm det} ( G_{\mu\nu} \partial_a X^\mu  \partial_b X^\nu)  } 
+\frac{i}{2} \epsilon^{ab} B_{\mu\nu}   \partial_a X^\mu \partial_b X^\nu 
\right). 
\end{equation}
The equations of motion are affected by the presence of the NS field, therefore the
minimal surface corresponding to the cusp Wilson line found by 
 \cite{Kruczenski:2002fb} will be modified. Note that we work with the  Euclidean world sheet 
 action. The same analysis can be performed in the Minkowski signature on the 
 world sheet with identical results as was done in the absence of 
 the NS-NS 3-form flux in \cite{Alday:2007hr}.  
 We look for a minimal surface with the following ansatz
 \begin{eqnarray}
\label{an2}
\Lambda z=e^{\t-\s} g(\s),\quad \Lambda x_0=e^{\t-\s} \cosh(\s+\t),\quad 
\Lambda x_1=e^{\t-\s} \sinh(\s+\t), 
\end{eqnarray}
where $\sigma, \tau$ are the world sheet co-ordinates and $\Lambda$ is an arbitrary 
scale. We will show that the action of the minimal surface is independent of the scale
$\Lambda$.  The ansatz in (\ref{an2}) satisfies the property 
that 
\begin{equation}
\frac{z^2}{ x_0^2 - x_1^2} = g^2 . 
\end{equation}
Therefore if $g$ does not vanish at any point in the co-ordinate $\sigma$, the surface 
reduces to a light cone at the boundary of $AdS_3$ at  $z=0$. 
The equations of motion for $g(\s)$ following from the Nambu-Goto action is given by 
\begin{eqnarray}\label{g0}
8 + 12 g g' + 3 g^2 \left[(g')^2-4\right]
-g^3 \left(g''+6 g' \right)+4 g^4 =
i 8 b  \left(-g g'+g^2-1\right)^{3/2} , 
\end{eqnarray}
where the primes refer to derivatives with respect to $\sigma$. 
The above differential equation admits a simple exact solution if we assume $g'=0$ and therefore
$g = c$ where $c$ is a constant. 
From (\ref{g0}) we see that $c$ then must satisfy the algebraic equation
\begin{equation}
c^2  - 2 = i 2 b \sqrt{ ( c^2 - 1) }. 
\end{equation}
The solutions to this equation  are given by 
\begin{eqnarray}\label{valc}
c &=& \sqrt{2} ( 1- b^2 \pm i b \sqrt{ 1- b^2} ) ^{\frac{1}{2}}, \\ \nonumber
& = & \sqrt{2}  \pm i \frac{b}{\sqrt{2}} - \frac{3 }{4\sqrt{2}} b^2 + \cdots. 
\end{eqnarray}
Here we have kept the roots which reduce to the solution found by \cite{Kruczenski:2002fb}
at $b=0$. Note that this is a complex solution to the equations of the motion when $b\neq 0$. 
We call this solution the `uniform' solution since $g(\sigma)$ does not depend on the 
world sheet co-ordinate $\sigma$.  
In general one can solve the equation of motion given in (\ref{g0}) perturbatively in 
$b$ for $g$ which is not uniform in the world sheet co-ordinate.
We will see that the spinning string solution in (\ref{scalingsol}) can be mapped to a non-uniform 
solution but in a further scaling limit it reduces to the uniform solution. 

Let us now evaluate the action of the `uniform' solution. 
We follow the regularization procedure adopted by \cite{Kruczenski:2002fb}. 
First we define the following world sheet coordinates. 
\be\label{rhoxi}
\rho= \t-\s ,\quad \xi = \t+\s. 
\ee
In these coordinates the induced metric on the world sheet is given by  
\be
ds^2=\frac{1}{c^2}\left( \frac{c^2-1}{\r^2}d\r^2+d\xi^2\right). 
\ee
Following \cite{Kruczenski:2002fb} we take 
the range for the coordinates $\rho, \xi$ to be  $(\epsilon, L) $ and $( -\gamma/2 , \gamma/2)$ respectively. 
$\gamma$ is then the cusp angle, $\epsilon, L$ are the UV and IR cutoff's respectively. 
Substituting the induced metric into the Nambu-Goto action (\ref{ngact}) we obtain
\begin{eqnarray}\label{sngw}
S_{NG}&=&\frac{\sqrt{\l}}{2\pi}\int_{-\g/2}^{\g/2} \int_{\e}^{L}
\left(\sqrt{G}+i\frac{b}{z^2}\left(x_0'\dot{x_1}-x_1'\dot{x_0}\right)\right)d\r d\xi \\ \nonumber
&=&\frac{\sqrt{\l}}{2\pi} \left[\frac{\sqrt{c^2-1}}{c^2}+i \frac{b}{c^2}\right] \g \log \frac{L}{\e}. 
\end{eqnarray}
Substituting the solution for $c$ from (\ref{valc}) and expanding in powers of $b$ we obtain
\begin{eqnarray}
S_{NG} &=& 
\frac{\sqrt{\l}}{2\pi} \left[\frac{ ( 1- 2 b^2 \pm 2i b \sqrt{1-b^2} )^{1/2}+ i b }{2 \left( 1- b^2  \pm i b \sqrt{ 1- b^2 } \right)}\right]
\g \log \frac{L}{\e}, \\ \nonumber
&=& \frac{\sqrt{\l}}{2\pi} \left[\frac{1}{2}\pm i\frac{b}{2}-\frac{b^2}{4}+O(\tb^3)\right]\g \log \frac{L}{\e}. 
\end{eqnarray}
Therefore the expectation value of the light like  Wilson loop is given by 
\begin{equation}\label{areawng}
 \langle W\rangle = \exp({-S_{NG} }) = 
 \left( \frac{L}{\epsilon}\right) ^{-\frac{\sqrt{\l}}{2\pi} \left[\frac{1}{2}\pm i \frac{b}{2}-\frac{b^2}{4}+O(\tb^3)\right]\g}. 
\end{equation}
The imaginary term just contributes to a phase. Note however the modulus 
of the expectation value or the area is corrected at $O(b^2)$. 
The coefficient of this correction is precisely $1/2$ of the leading term and of the opposite sign. 
This is precisely the behaviour of the coefficient of leading correction  at $O(b^2 \log^2 S) $ of the anomalous dimension 
seen in (\ref{Disp2}). 
Though this analysis has been done in Euclidean world sheet  signature, it can be repeated 
with Minkowski  world sheet signature to arrive at the same conclusion as 
done for the $b=0$ case in \cite{Alday:2007hr}.

\subsection{Relating the Wilson loop and the spinning string} \label{relatewilspin}

We have observed that the $O(b^2)$ correction of the cusp anomaly   precsiely agrees with the 
$O(b^2 \log^2 S) $ term in the anomalous dimension of the spinning string. 
Just as the $O(b^0)$ of the cusp anomaly agrees with the $O(\log S)$ term in the dispersion 
relation of the spinning string. 
To relate these solutions further we follow the method of \cite{Kruczenski:2007cy} and perform a set of 
conformal transformation and re-parametrizations starting from the scaling limit of the spinning 
string solution in (\ref{scalingsol}) and arrive at the light like Wilson loop. 
We will see that after a further scaling limit, the solution is precisely that of the 
`uniform' Wilson loop. 

To begin let us define the embedding co-ordinates in which 
$AdS_3$ is a hyperboloid and its relationship with the $AdS_3$ global co-ordinate in which 
the solution (\ref{scalingsol}) is written down. 
The hyperboloid   is defined by the constraint  
\begin{equation} \label{hyperboloid}
 -X_0^2 - X_3^2 + X_1^2 + X_2^2 = -1. 
\end{equation}
and the metric in the embedding space is given by 
\begin{equation}
 ds^2 = - dX_0^2 - dX_3^2 + dX_1^2 + dX_2^2 . 
\end{equation}
The relationship between these co-ordinates and the global co-ordinates of $AdS_3$ given in (\ref{metrics}) is given by 
\begin{eqnarray}\label{poin1}
X_0&=&\cosh \rho \cos t ,\quad X_3=\cosh \rho \sin t,\\ \nn
  X_1&=&\sinh \rho \cos \phi, \quad X_2=\sinh \rho \sin \phi.  \nn
\end{eqnarray}

We will now outline the transformations  to relate the spinning solution (\ref{scalingsol}) 
to the light-like Wilson loop. 
\begin{enumerate}
 \item  We first analytically continue the Minkowski world sheet to Euclidean by replacing 
 \begin{equation}
  \tau = i \tilde \tau. 
 \end{equation}
 For convenience we also replace 
 \begin{equation}
  b = i \tilde b. 
 \end{equation}
This is so that we do not have to deal with intermediate factors of $i$. At the end of all the steps
we will reinstate $b$. 
 On performing this we see that 
 the solution in the embedding co-ordinates is of the form
\begin{eqnarray} \label{embed1}
X_0&=&\cosh \rho \cosh \tilde{t} ,\quad X_3=i \cosh \rho \sinh\tilde{ t}, \\ \nn
\quad X_1&=&\sinh \rho \cosh \tilde{\phi}, \quad X_2=i \sinh \rho \sinh \tilde{\phi}.  
\end{eqnarray}
where $\rho, \tilde t, \tilde\phi$ are the solutions given in (\ref{scalingsol}) but with $\tau, b$ replaced with 
$\tilde \tau, \tilde b$ respectively. 
\item The next step is to factor out the pre-factor  $i$  in    $X_3$ and $X_2$ in (\ref{embed1}) and then 
exchange  $3\rightarrow 2$. It is clear from that this operation still preserves the 
constraint (\ref{hyperboloid}). Therefore we obtain the solution
\begin{eqnarray}
X_0'&=&\cosh \rho \cosh \tilde{t} ,\quad X_2'= \cosh \rho \sinh\tilde{ t}, \\ \nn
\quad X_1'&=&\sinh \rho \cosh \tilde{\phi}, \quad X_3'= \sinh \rho \sinh \tilde{\phi}. 
\end{eqnarray}
\item We now perform $2$ rotations in the $0-3$ plane and $2-1$ plane, each with  angle $\pi/4$. 
Therefore we obtain the solution
\begin{eqnarray} \label{wilemb}
X_{0}^{\rm w }=\frac{X_0'+X_3'}{\sqrt{2}},\quad X_{3}^{\rm w }=\frac{X_0'-X_3'}{\sqrt{2}}, \\ \nn
X_{1}^{\rm w }=\frac{X_2'+X_1'}{\sqrt{2}},\quad X_{2}^{\rm w}=\frac{X_2'-X_1'}{\sqrt{2}}. 
\end{eqnarray}
This solution is the Wilson loop in the embedding co-ordinates. 
\item
Finally we write the solution (\ref{wilemb}) in  the Poincar\'{e} patch by using the 
following relations
\be
\label{emb2poin}
X_3^{\rm w} -X_2^{\rm w}=\frac{1}{\Lambda z},\quad 
X_0^{\rm w} =\frac{x_0}{ z},\quad 
X_1^{\rm w} =\frac{x_1}{z},\quad X_3^{\rm w} +X_2^{\rm w} = \Lambda ( z+\frac{x^2}{z}  ) . 
\ee
where $x^2=-x_0^2+x_1^2$.  Recall that $z, x^0, x^1$ are the co-ordinates in the
Poincar\'{e} patch with metric given in (\ref{poinmet}). 
\end{enumerate}

After performing these operations on the spinning 
string solution given in (\ref{scalingsol})  we obtain the following solution in the Poincar\'{e} patch. 
\begin{eqnarray}\label{wilsol0}
\Lambda z&=&\sqrt{2} e^{\tau -\sigma }+\frac{\tb}{\sqrt{2}} \left(e^{2 \sigma } (2 \sigma -\pi  c_{1}^{(0)})+2 \pi  
c_{1}^{(0)}-2\right) e^{\tau -3 \sigma } \\ \nonumber
& & +\frac{\tb^2 e^{\tau -3 \sigma }}{4 \sqrt{2}}\Big\{4 (1-\pi  c_{1}^{(0)}) (\pi  c_{1}^{(0)}-2 \sigma +2) \\ \nn
&&+ \cosh (2 \sigma ) \left(11 \pi ^2 (c_{1}^{(0)})^2+4 \pi  c_{1}^{(0)} (-2 \sigma +\tau -5)+4 (\sigma^2 - \sigma +3)\right)\\ \nn
&&+\sinh (2 \sigma ) \left(-5 \pi ^2 (c_{1}^{(0)})^2+4 \pi  c_{1}^{(0)} (-2 \sigma +\tau +3)+4 (\sigma^2 - \sigma -1)\right)\Big\}, \\ \nn
\Lambda x_0&=&
e^{\t-\s}\cosh(\t+\s)+\frac{1}{2} \tb e^{-4 \sigma } \left\{
e^{2\tau + 4\sigma} (2 \sigma -\pi  c_1^{(0)})+ e^{2\tau + 2\sigma} ( 2 \pi  c_1^{(0)}-2 )
-e^{2\sigma} +\pi  c_1^{(0)}-1\right\}\\ \nonumber
&&+\frac{1}{4} \tb^2 e^{-6 \sigma } \Big\{4 (\pi  c_1^{(0)}-1)^2 e^{2 (\sigma +\tau )}+4 (\pi  c_1^{(0)}-1) (2 \sigma -\pi  c_1^{(0)}) e^{4 \sigma +2 \tau } \\ \nonumber & & +(4-4 \pi  c_1^{(0)}) e^{2 \sigma }  
  +2 (\pi  c_1^{(0)}-1)^2 
+e^{6 \sigma +2 \tau } \left((\pi  c_1^{(0)}-2 \sigma )^2 +2 \pi  c_1^{(0)} \tau \right) (\pi  c_1^{(0)}-1)^2
\\ \nonumber
& & +e^{4 \sigma } (\pi  c_1^{(0)} (\pi  c_1^{(0)}-2 \sigma -2)-2 \sigma +2)\Big\}, \\ \nn
\Lambda x_1&=&e^{\t-\s}\sinh(\t+\s)+\frac{1}{2} \tb e^{-4 \sigma } \left\{ 
e^{2\tau + 4 \sigma } (2 \sigma -\pi  c_1^{(0)})+ e^{2\tau + 2\sigma}( 2 \pi  c_1^{(0)}-2)  + e^{2\sigma} 
-\pi  c_1^{(0)} +1\right\} \\ \nonumber
&&+ \frac{1}{4} \tb^2 e^{-6 \sigma } \Big(4 (\pi  c_1^{(0)}-1)^2 e^{2 (\sigma +\tau )}+4 (\pi  c_1^{(0)}-1) (2 \sigma -\pi  c_1^{(0)}) e^{4 \sigma +2 \tau } \\ \nonumber
& & +4 (\pi  c_1^{(0)}-1) e^{2 \sigma }-2 (\pi  c_1^{(0)}-1)^2
+e^{6 \sigma +2 \tau } \left((\pi  c_1^{(0)}-2 \sigma )^2+2 \pi  c_1^{(0)} \tau \right) \\ \nonumber
& & +e^{4 \sigma } (\pi  c_1^{(0)} (-\pi  c_1^{(0)}+2 \sigma +2)+2 (\sigma -1))\Big).  \nonumber
\end{eqnarray}
One can explicitly verify  that the  expressions for $x$, $x_0$ and $x_1$,  given in 
(\ref{wilsol0})  are solutions to the equations of motion of the  Nambu-Goto action (\ref{ngact})
to order $b^2$. 

The solution given in (\ref{wilsol0}) seems complicated, but we can see that it 
ends on a 
a light like cusp at the $AdS_3$ boundary. To  show this  we compute
\be
\frac{z^2}{x_0^2-x_1^2}=G=2 \tb^2 e^{-4 \sigma } \left(-\pi  c_1^{(0)}+e^{2 \sigma }+1\right)^2+\tb \left((2-2 \pi  c_1^{(0)}) e^{-2 \sigma }+2\right)+2. 
\ee
Since $b<<1$, the quantity $G$ does not vanish \footnote{ For $G$ to vanish it can be 
shown that $e^{2\s}$ has to become  complex.}, therefore at $z=0$, the boundary of 
$AdS$,  the minimal surface in (\ref{wilsol0}) ends on the light cone
$x_0=\pm x_1$. 
To make the resemblance with the minimal surface corresponding to the 
light like cusp more apparent we can perform a world sheet re-parametrization.
Let us first equate  the solution in (\ref{wilsol0}) to the  following general ansatz. 
\be
\Lambda z=f(\t,\s) g(\t,\s),\quad  \Lambda x_0=f(\t,\s)\cosh(X(\t,\s)),\quad \Lambda x_1=f(\t,\s)\sinh(X(\t,\s)). 
\ee
After a bit of tedious 
 algebra we can extract the functions $f(\t,\s)$, $g(\t,\s)$ and $X(\t,\s)$ to order $b^2$.
 These are given by 
\begin{eqnarray}
f(\t,\s)&=&e^{\tau -\sigma }+\frac{1}{2} \tb \left(e^{2 \sigma } (-\pi  c_1^{(0)}+2 \sigma -1)+3 \pi  c_1^{(0)}-3\right) e^{\tau -3 \sigma }+\\ \nonumber 
&&\frac{1}{8} \tb^2 e^{\tau -5 \sigma } \Big\{2 (\pi  c_1^{(0)}-1) e^{2 \sigma } (-3 \pi  c_1^{(0)}+6 \sigma -5)+11 (\pi  c_1^{(0)}-1)^2+ \\ \nonumber
&&e^{4 \sigma } \left(3 \pi ^2 (c_1^{(0)})^2-2 \pi  c_1^{(0)} (4 \sigma -2 \tau +1)+4 (\sigma -2) \sigma +3\right)\Big\}, \\ \nn
g(\t,\s)&=&\sqrt{2}+\frac{\tb }{\sqrt{2}}e^{-2 \sigma } \left(-\pi  c_1^{(0)}+e^{2 \sigma }+1\right)+
\frac{3 \tb^2 }{4 \sqrt{2}}e^{-4 \sigma } \left(-\pi  c_1^{(0)}+e^{2 \sigma }+1\right)^2, \\ \nonumber
X(\t,\s)&=&\sigma +\tau + \tb e^{-\sigma } \left\{(-\pi  c_1^{(0)}+\sigma +1) \sinh (\sigma )+\sigma  \cosh (\sigma )\right\}
+\frac{1}{2} \tb^2 e^{-2 \sigma } \left\{\pi  c_1^{(0)}-1\right. \\ \nonumber
&&\left. + \sinh (2 \sigma ) (\pi  c_1^{(0)} \sigma +\pi  c_1^{(0)} \tau +\sigma )+\cosh (2 \sigma ) \left[\pi  c_1^{(0)} (-\pi  c_1^{(0)}+\sigma +\tau +2)+\sigma -1\right]\right\}. 
\end{eqnarray}
Then we re-parametrize the world sheet co-ordinates  by introducing 
co-ordinates  which satisfy the condition
\be
\t'+\s'=X(\t,\s), \quad \t'-\s'=\log[f(\t,\s)].
\ee
From these equations it is possible express the co-ordinates $\tau, \sigma$ in terms
$\tau', \sigma'$ perturbatively in $b$. This change of variables is given by 
\begin{eqnarray}
\t&=& \t'+\tb \left\{ (1-\pi  c_1^{(0)}) e^{-2 \s'}+\frac{\pi  c_1^{(0)}}{2}-\s'\right\}+\\ \nn
&& \frac{1}{2} \tb^2 
\left\{2 (\pi c_1^{(0)}-1)^2 e^{-4 \s'}+(3-3 \pi c_1^{(0)}) e^{-2 \s'}-\pi c_1^{(0)} \t'+1\right\},  \\ \nonumber
\s&=&\s'-\frac{1}{2} \tb e^{-2 \s'} \left(-\pi c_1^{(0)}+e^{2 \s'}+1\right) \\  \nonumber
& & +\frac{1}{4} \tb^2 e^{-4 \s'} \left\{ e^{4 \s'} \left((\pi c_1^{(0)}-1)^2-2 (\pi c_1^{(0)}+1) \s'\right)-(\pi c_1^{(0)}-1)^2\right\}. 
\end{eqnarray}
After this reparametrization, we can write the solutions as
\be\label{wilsol1}
\Lambda z=e^{\t'-\s'} g(\t',\s'),\quad \Lambda x_0=e^{\t'-\s'}\cosh(\t'+\s'),\quad 
\Lambda x_1=e^{\t'-\s'}\sinh(\t'+\s'), 
\ee
where
\begin{eqnarray}\label{g}
g(\t',\s')&=&\sqrt{2}+\frac{\tb }{\sqrt{2}}e^{-2 \s'} \left(-\pi c_1^{(0)}+e^{2 \s'}+1\right)+\\ \nn
&&\frac{\tb^2 }{4 \sqrt{2}}e^{-4 \s'} \left(-7 \pi c_1^{(0)}+3 e^{2 \s'}+7\right) \left(-\pi c_1^{(0)}+e^{2 \s'}+1\right).  \\ \nn
\end{eqnarray}
To verify  all the manipulations performed we have checked that the minimal surface 
given in (\ref{wilsol1}) solves the Nambu-Goto equations of motion. 
Since the ansatz in (\ref{an2}) is of the same form given in (\ref{wilsol1}), it can be verified that
the expression for $g$ given in (\ref{g}) solves the equations of motion 
(\ref{g0}) to order $b^2$.  Note that the solution given in (\ref{wilsol1}) and (\ref{g})
is a solution for any arbitrary $c_1^{(0)}$ and it is interesting to note that it reduces
to the `uniform' solution  $g=c$ with $c$ given in (\ref{valc})   for 
$c_1^{(0)}=\frac{1}{\pi}$. Therefore this solution is a one parameter generalization of the 
`uniform' solution. 

Let us now  go over to the $\rho$ and $\xi$ co-ordinates introduced in (\ref{rhoxi}) to 
evaluate the Euclidean action of the solution. 
We define
\be
\t'=\frac{1}{2} (\xi +\log (\rho )),\quad \s'=\frac{1}{2} (\xi -\log (\rho )), 
\ee
in terms of these co-ordinates,  the solutions become
\begin{eqnarray}\label{finalwilson}
\Lambda z&=&\sqrt{2}\rho
+\frac{\tb}{\sqrt{2}}  e^{-\xi } \rho  \left(-\pi  c_1^{(0)} \rho +e^{\xi }+\rho \right) +\\ \nonumber
&&\frac{\tb^2 }{4 \sqrt{2}}
e^{-2 \xi } \rho  \left(-7 \pi c_1^{(0)} \rho +3 e^{\xi }+7 \rho \right) \left(-\pi c_1^{(0)} \rho +e^{\xi }+\rho \right), 
\\ \nonumber
\Lambda x_0&=&\rho \cosh (\xi), \quad \Lambda x_1=\rho \sinh (\xi).
\end{eqnarray}
We now further scale 
\begin{equation}\label{newscal}
\rho = \Lambda \rho', \qquad  \Lambda \rightarrow 0, \quad \rho':\; \; \mbox{finite}. 
\end{equation}
Then it is easy to see that solution for $z$ reduces to  $z = c\rho'$ where $c$ is given in 
(\ref{valc}) and $x_0 = \rho' \cosh(\xi),  x_1 = \rho' \sinh(\xi)$. 
Thus the Euclidean action of the solution in this  scaling limit will be given by 
(\ref{areawng}). 
Therefore the coefficient of the $b^2$ term in the  area of the Wilson surface 
obtained from  the spinning string solution by conformal transformations and 
re-parametrizations exhibits the same behaviour as the $O(b^2 (\log S)^2)$ term
in the dispersion relation of the spinning string.

One might wonder what would be the result if one were not to perform the scaling in 
(\ref{newscal}). We have verified that the coefficient of the log divergence in the
area of the Wilson loop  still remains the same. This can be seen by 
evaluating  Nambu-Goto action for the solution given in (\ref{finalwilson})
\begin{eqnarray}
S_{NG}&=&\frac{\sqrt{\l}}{2\pi}\int_{-\g/2}^{\g/2} \int_{\e}^{L}
\left(\sqrt{G}+\frac{\tb}{z^2}\left(x_0'\dot{x_1}-x_1'\dot{x_0}\right)\right)d\r d\xi \\ \nonumber
&=&\frac{\sqrt{\l}}{2\pi}\Bigg[\frac{1}{2}\log \frac{L}{\e}+\tb \left(\frac{\g}{2}\log \frac{L}{\e}-(c_1^{(0)}\pi-1)(L-\e)\sinh \frac{\g}{2}\right)+\\ \nn
&& \frac{\tb^2}{4} \left(\gamma  \log \frac{L}{\e}+(\pi  c_1^{(0)}-1) \left((\pi  c_1^{(0)}-1) \sinh (\gamma ) (L^2-\epsilon^2 )+4 \sinh \left(\frac{\gamma }{2}\right) (\epsilon -L)\right)\right)\Bigg]
\\ 
&&+O(\tb^3)\nn
\end{eqnarray}
where $\dot{x_i}$ and $x_i'$ represent derivative with respect to $\rho$ and $\xi$ respectively.
We have to reinstate $\tb=-i b$ to obtain the answer. Thus there are
are non-universal  quadratic divergent terms  in the 
area that depends on the parameter $c_1^{(0)}$. 
The coefficients of these terms vary if one changes the integration constants in
(\ref{scalingsol}), but the coefficient of the log-divergence is invariant and universal.
This observation together with the fact that the classical solution of the 
spinning string is related to the cusp minimal surface suggests that 
the  $O(b^2)$  coefficient of logarithmic divergence of the area of the Wilson loop is 
related to the coefficient of $O(b^2 \log^2 {\cal S})$ in the 
dispersion relation of the spinning string. 

\section{Conclusions}

We have studied classical spinning strings and their dispersion relation  in the
$AdS_3$ with mixed 3-form fluxes. 
We have shown that the dispersion relation acquires the 
term $-\frac{\sqrt{\lambda} b^2}{2\pi} \log^2 {\cal S}$ in addition to the usual $\log {\cal S}$ term. 
We have observed that the the coefficient of the $b^2$ term in the logarithmic divergence
of the area of the minimal surface corresponding to the  cusp-Wilson line 
is identical to the correction in the dispersion relation of the folded spinning string.
This observation together with the fact that the spinning string in the presence of the 
NS-flux can be mapped to the minimal surface suggests that 
the coefficient of this term can be derived to all orders in the 
coupling $\lambda$.  It will be interesting to study this observation further.

There has been progress in writing down the S-matirx of strings in 
$AdS_3\times S^3\times {\cal M}$
\cite{Babichenko:2009dk,David:2010yg,OhlssonSax:2011ms,Sax:2012jv,Borsato:2012ud,Borsato:2012ss,Borsato:2013qpa,Borsato:2013hoa,Lloyd:2013wza,Borsato:2014exa}.
This has been extended to the case with mixed 3-form fluxes in 
\cite{Hoare:2013pma,Hoare:2013ida,Bianchi:2014rfa}. 
The results obtained in this paper will serve as tests of these
proposals. For the 
case of ${\cal N}=4$ Yang-Mills a crucial step in understanding of the 
S-matrix was the derivation of the cusp-anomalous dimension to all 
order in the coupling \cite{Beisert:2006ez}. 
Our results suggest that the
cusp-anomalous dimension 
in $AdS_3\times S^3$ has an interesting 
deformation parametrized by the NS-B flux. 
Deriving the cusp anomalous dimension from the S-matrix 
to all orders in the coupling  as well its deformation 
in the NS-B flux is an important direction to pursue in this 
subject. It will  lead to crucial insights for the strings in $AdS_3\times S^3$ and its
holographic dual. 

{\bf Note added}: While this paper was being written up we noticed \cite{Ahn:2014tua}
on the arXiv where classical strings in $AdS_3\times S^3$ was studied with
an emphasis on the giant magnon solution.

\acknowledgments
We wish to thank  Dileep Jatkar, Rajesh Gopakumar and Ashoke Sen for Discussions.
A.S wishes to thank string group at 
 Harish-Chandra Research Institute, Allahabad for hospitality 
during part of this project. 
J.R.D wishes to thank the Simons Centre for Geometry and Physics, Stony Brook and the
organizers of the ``Quantum Anomalies, Topology and Hydrodynamics'' program  
for hospitality 
during the completion of the project. 
The work of J.R.D is partially supported by the Ramanujan fellowship DST-SR/S2/RJN-59/2009, the work of A.S is supported by a CSIR fellowship (File no: 09/079(2372)/2010-EMR-I).

\bibliographystyle{JHEP}
 \bibliography{ads3}

\end{document}